\def\sect{\frenchspacing Section }

\def\fig{\frenchspacing Fig. }
\def\tab{\frenchspacing Table }
   
\documentclass[usenatbib]{mn2e} 
\usepackage{graphicx,epsfig}
\usepackage{amssymb}

\title{The study of topology of the universe using multipole vectors}

\author[P.~Bielewicz, A.~Riazuelo]{P.~Bielewicz$^1$ \thanks{E-mail:
bielewic@iap.fr}, A.~Riazuelo$^1$\\ $^{1}$ Insitute d'Astrophysique de
Paris, 98bis boulevard Arago, 75014, Paris, France \\ }

\begin{document}

\maketitle

\begin{abstract}
We study a multipole vector-based decomposition of cosmic microwave
background (CMB) data in order to search for signatures of a multiconnected
topology of the universe. Using $10^6$ simulated maps, we analyse the
multipole vector distribution on the sky for the lowest order multipoles
together with the probability distribution function of statistics based
on the sum of the dot products of the multipole vectors for both the
simply-connected flat universe and universes with the topology of a
3-torus. The estimated probabilities of obtaining lower values for these
statistics as compared to the 5-year \emph{WMAP} data indicate
that the observed alignment of the quadrupole and octopole is
statistically favoured in a 3-torus topology where at least one
dimension of the fundamental domain is significantly shorter than the
diameter of the observable universe, as compared to the usual standard
simply-connected universe. However, none of the obtained results are able to
clearly rule out the latter (at more than 97\% confidence
level). Multipole vector statistics do not appear
to be very sensitive to the signatures of a 3-torus topology if the
shorter dimension of the domain becomes comparable to the diameter of the
observable universe. Unfortunately, the signatures are also significantly
diluted by the integrated Sachs-Wolfe effect.
\end{abstract}


\section{Introduction}

According to General Relativity, a pseudo-Riemannian manifold with
signature (3,1) is the mathematical model of spacetime. The local
properties of spacetime geometry are described by the Einstein
gravitational field equations. However, they are only weakly linked
\citep{roukema:2007} to the global spatial geometry of the universe,
i.e., its topology. We do not have, so far, any theory which could
determine the topology of our universe, therefore it can be constrained
only by observations.

The release of \emph{WMAP} data
\citep{bennett:2003,hinshaw:2007,hinshaw:2009} has stimulated new
studies of multiconnected (incorrectly dubbed as ``non-trivial'')
topology recently, particularly since the detected anomalies in the
observed cosmic microwave background (CMB) anisotropy on large angular
scales --
the putative suppression of the quadrupole moment, alignment of the
quadrupole and octopole, and an asymmetry in the statistical properties
of the northern and southern ecliptic hemispheres 
\citep{de Oliveira-Costa:2004, copi:2004, eriksen:2004,
hansen:2004, schwarz:2004} -- suggest that our universe may possess
multiconnected topology.

Amongst all of the multi-connected three-dimensional spaces, flat spaces, and
the most natural compact manifold, the 3-torus, have been
studied the most extensively in a cosmological context. This is
motivated by current constraints on the curvature radius of the universe as
well as computational simplicity.  One of the first related analyses
\citep{tegmark:2003} suggested that a
toroidal universe where the smaller dimension was of order half the
horizon scale may explain the above anomalies, although a more detailed study
\citep{de Oliveira-Costa:2004} did not confirm this hypothesis.  Another
topology, the Poincar\'e dodecahedron for slightly positively curved
space, was proposed by \cite{luminet:2003}. Further studies of this
topological model revealed some disagreement about whether the topology
could be excluded on the basis of the \emph{WMAP} data or not. 
\cite{aurich:2005a,aurich:2005b,aurich:2006,caillerie:2007} claimed that
the topology can not be excluded, mainly because of insufficient
accuracy of the data and degradation of the signal by the integrated
Sachs-Wolfe (ISW) and Doppler contributions. Moreover, \cite{roukema:2004}
even reported a hint of a detection of pairs of matched circles in the CMB
maps which fit quite well to the predictions for this topology.
However, \cite{cornish:2004} and \cite{key:2007} ruled it
out via the lack of detection of statistically significant pairs of
matched circles \citep[see also][]{lew:2008,roukema:2008}.

Topological studies of CMB maps generally focus on two signatures 
of multi-connectedness: the large scale damping of power in the
direction of the shorter dimension of the domain, which causes a 
breakdown of statistical isotropy
\citep{hajian:2003,kunz:2006,kunz:2008,niarchou:2006,niarchou:2007}, and
the distribution of matched patterns
\citep{levin:1998,cornish:1998,bond:2000a,bond:2000b,cornish:2004,roukema:2004,
aurich:2005a,aurich:2005b,key:2007} \cite{levin:2002} presents a review of
the CMB-related methods used in studies of topology.  It is important to notice
that the latter signature is present only when the topological scale is
smaller than observable universe, thus methods based on it are less
powerful than methods based on the former signatures of
multi-connectedness, which are in principle present also if the
topological scale is slightly greater than the particle horizon.

In the former case, one usually uses the covariance matrix of the spherical
harmonic coefficients \citep{kunz:2006,kunz:2008} of the sky map 
as the basis of a statistical study.
However, \cite{aurich:2007} used also the so-called multipole vectors, 
initially proposed by \cite{copi:2004} as a new tool for the 
analysis of CMB maps. In contrast to
the coordinate dependent coefficients of the spherical harmonic
decomposition, the vectors associated with a given multipole point toward the same
direction on the sky independently of the reference frame employed. Thus,
they are useful tools for the study of statistical isotropy
(irrespective of its topological origin, as we consider here, or otherwise). 
In this paper, we
study in more detail the application of the
multipole vector formalism to the detection of signatures of multiconnected 
topology in the context of the observed alignment of the quadrupole and octopole
moments. Because our studies have a rather preliminary character, we will
confine ourselves in analysis to the simplest compact
topology, i.e., 3-torus that remains to be ruled out by the data
\citep{aurich:2008}.

In the following two sections we briefly introduce the multipole vectors
and related statistics used in the analysis. In \sect\ref{sec:topology} we describe the
specific topologies studied. The results of the analysis 
comparing the 5-year \emph{WMAP} data to the
studied topologies are presented in \sect\ref{sec:results}. Finally,
we summarise the results and draw some conclusions.

\section{Multipole vector decomposition}

The multipole vector formalism was introduced to CMB analysis by
\citet{copi:2004}, who showed that a given multipole moment can be
represented in terms of $\ell$ unit vectors and an overall magnitude,
thus making as expected $2 \ell + 1$ figures. As later pointed out by
\citet{weeks:2004}, the formalism was in fact first discovered long ago
by \citet{maxwell:1891}. Each multipole $T_\ell$ may therefore be
uniquely expressed by $\ell$ multipole vectors
$\hat{\mathbf{v}}^{(\ell,1)}, \ldots, \hat{\mathbf{v}}^{(\ell,\ell)}$
and a magnitude $A^{(\ell)}$. In the notation of \citet{copi:2004}, this
reads
\begin{equation} 
T_\ell(\hat{\mathbf{e}})
 \equiv \sum_{m=-\ell}^\ell a_{\ell m} Y_{\ell
m}(\hat{\mathbf{e}})
 = A^{(\ell)} (\hat{\mathbf{v}}^{(\ell,1)} \cdot \hat{\mathbf{e}}) \dots
(\hat{\mathbf{v}}^{(\ell,\ell)} \cdot \hat{\mathbf{e}}) + Q \ ,
\end{equation}
where $\hat{\mathbf{e}}$ is the radial unit vector in spherical
coordinates. Strictly speaking, the multipole vectors are headless, thus
the sign of each vector can always be absorbed by the scalar
$A^{(\ell)}$. We will use the convention that all vectors point toward
the northern hemisphere.

It is hard to find any convincing physical interpretation of the multipole
vectors, because they do not indicate any specific features of the
multipoles. An exception to this rule arises when one considers the
contribution to a multipole given by the projection on the last
scattering surface of a single plane wave. In this case, all the
multipole vectors point towards the direction of the wavevector
\citep{mlr:2004}. Apart from this very specific case, for
the quadrupole one can notice that the multipole vectors point towards the
inflection points, such that their cross product points at the saddle
points. For the higher order multipoles,
the vectors indicate at vicinity of the inflection points rather
than extremes.

To find the multipole vectors from a given set of spherical harmonic
coefficients, $a_{\ell m}$, one has to solve a set of non-linear equations
\citep{copi:2004}. For a quadrupole the solution of the equations can be
written explicitly (see appendix \ref{sec:appendix}). One should notice that
the multipole vectors are related non-linearly to the data. For higher
order multipoles the equations are solved numerically. In our codes we
have implemented the algorithm proposed\footnote{The routines of
\citet{copi:2004} are available at http URL
http://www.phys.cwru.edu/projects/mpvectors/~.} by
\citet{copi:2004}. An alternative was suggested by
\citet{katz:2004}.

\section{Multipole vectors statistics}

Having computed the multipole vectors, we seek to test a given CMB data
set with respect to inter-scale correlations between different
multipoles. In order to do so, we follow \citet{copi:2004} and
\citet{schwarz:2004} and define the following set of simple statistics.

The statistics are based on the dot product, which is a natural measure
of vector alignment. However, because the multipole vectors are defined
only up to a sign, one needs to consider the absolute value of the dot
product.  Moreover, the multipole vectors do not have their own
identity, therefore all the dot products were summed for a given pair of
the multipoles $\ell_1$ and $\ell_2$. At the end, to normalise the
statistics to one, the sum were divided by number of the dot products.
We consider also the dot products of the unnormalised cross products of
the multipole vectors, $\mathbf{w}^{(\ell,i)} =
\hat{\mathbf{v}}^{(\ell,j)} \times \hat{\mathbf{v}}^{(\ell,k)}$, as well
normalised cross products, $\hat{\mathbf{w}}^{(\ell,i)} =
(\hat{\mathbf{v}}^{(\ell,j)} \times \hat{\mathbf{v}}^{(\ell,k)}) / |
\hat{\mathbf{v}}^{(\ell,j)} \times \hat{\mathbf{v}}^{(\ell,k)} |$ (where
$j \neq k,\ j,k = 1,\dots, \ell$ and $i =
1,\dots,\ell(\ell-1)/2$). Therefore, we could use the following types of
statistics for any two multipoles $\ell_1$ and $\ell_2$ ($\ell_1 \neq
\ell_2$):
\begin{eqnarray}
S_{\textrm{vv}} &=& {1\over M_{\rm vv}} \sum_{i=1}^{\ell_1} \sum_{j=1}^{\ell_2}
|\hat{\mathbf{v}}^{(\ell_1,i)}\cdot \hat{\mathbf{v}}^{(\ell_2,j)}| \ , \\
S_{\textrm{vw}} &=& {1\over M_{\rm vw}} \sum_{i=1}^{\ell_1}
\sum_{j=1}^{\ell_2(\ell_2-1)/2} | \hat{\mathbf{v}}^{(\ell_1,i)}\cdot
\hat{\mathbf{w}}^{(\ell_2,j)}| \ ,  \\
S_{\textrm{wv}} &=& {1\over M_{\rm wv}} \sum_{i=1}^{\ell_1(\ell_1-1)/2} 
\sum_{j=1}^{\ell_2} | \hat{\mathbf{w}}^{(\ell_1,i)}\cdot \hat{\mathbf{v}}^{(\ell_2,j)}| \ ,  \\
S_{\textrm{ww}} &=& {1\over M_{\rm ww}} \sum_{i=1}^{\ell_1(\ell_1-1)/2}
\sum_{j=1}^{\ell_2(\ell_2-1)/2} |\hat{\mathbf{w}}^{(\ell_1,i)}\cdot
\hat{\mathbf{w}}^{(\ell_2,j)}| \ , 
\end{eqnarray} 
where for the last three types of statistics we used also unnormalised
cross products containing more information than the normalised
products. Then, the statistics will be denoted by $S_{\textrm{vw}}^u$,
$S_{\textrm{wv}}^u$ and $S_{\textrm{ww}}^u$, respectively.  The four
types of the statistics are as usual referred to as ``vector-vector'',
``vector-cross'', ``cross-vector'' and ``cross-cross'' statistics,
respectively.  $M$ is number of dot products used for a given statistic
($M_{\rm vv} = \ell_1 \ell_2$, $M_{\rm vw} = \ell_1 \ell_2
(\ell_2-1)/2$, $M_{\rm wv} = \ell_2 \ell_1 (\ell_1 -1)/2$, $M_{\rm ww} =
\ell_1 (\ell_1-1) \ell_2 (\ell_2-1)/4$, respectively).  With such
normalisation the statistics take values in the range $[0,1]$.

Because the statistics are based on the dot product, they are
rotationally invariant and are sensitive only to the relative orientation of
the multipole vectors on the sky, and not on the absolute orientation of
the fundamental domain with respect to a given reference frame.

\section{Topology of universe} \label{sec:topology}

A detailed description of all possible topologies of three-dimensional
manifolds with constant curvature were given by \cite{wolf:1967} and
\cite{inoue:2001}. A description and summary of the topologies for
flat universes, which we will consider in this paper, and methods of
simulations of CMB maps for such universes were also given by
\cite{riazuelo:2004a} and \cite{riazuelo:2004b}.  We will briefly
describe here, following their formalism, basic information concerning
the topology of the 3-torus.

All flat spaces are obtained as the quotient $\mathbf{E}^3/\Gamma$ of
the three-dimensional Euclidean space $\mathbf{E}^3$ by a group $\Gamma$
of symmetries of $\mathbf{E}^3$ that is discrete and fixed-point free.
There are 18 such spaces among which is the standard Euclidean space
with trivial topology.  Among the 17 remaining ones, the ten compact flat
spaces are quotients of the 3-torus: six are orientable and the rest are
non-orientable.  The 3-torus is the quotient of Euclidean space under
the action of three linearly independent translations $\mathbf{T}_1$,
$\mathbf{T}_2$ and $\mathbf{T}_3$. The rectangular 3-torus is generated
by mutually orthogonal translations $\mathbf{T}_1 = (L_x,0,0)$,
$\mathbf{T}_2 = (0,L_y,0)$ and $\mathbf{T}_3 = (0,0,L_z)$ where the
allowed wave vectors of the cosmological perturbations $\mathbf{k}$
takes the form $\mathbf{k} = 2\pi\, (n_x/L_x, n_y/L_y, n_z/L_z)$, where
$n_x, n_y, n_z \in \mathbb{Z}$ and $L_x,\ L_y,\ L_z$ are dimensions of the
fundamental cell, which in this context is called the fundamental
domain.

Therefore, there are two effects of the multiconnected topology as compared
to the Euclidean space: the non-isotropic distribution of modes and
the discrete power spectrum of the curvature perturbations.  Both of these
will manifest themselves in the CMB maps mainly by the ordinary
Sachs-Wolfe effect, which dominates on large angular scales.  The
two other contributions to the maps, the Doppler and integrated
Sachs-Wolfe (ISW) effects, will rather dilute the signatures of the
topology. The former is significant on scales smaller than the dimensions
of the fundamental domain and the latter, though significant on
large angular scales, arises from the evolution of structure close to the
observer.

To see pronounced signatures of the topology we considered domains with
at least one of the dimensions smaller than the diameter of the observed
universe.  The distance to the horizon is approximately equal to the
distance to the last scattering surface (LSS) $\eta_{LSS}$, therefore
the diameter is $D \approx 2\, \eta_{LSS}$, but depends on the parameters of
the cosmological model. For the concordance $\Lambda$CDM model, used in
our work, with the Hubble constant \mbox{$H_0 = 70\ \rm{km}\,
\rm{s}^{-1} \rm{Mpc}^{-1}$}, the matter density parameter $\Omega_m =
0.3$, the cosmological constant density parameter $\Omega_\Lambda =
0.7$, the diameter is $D = 6.36\, R_H$, where the Hubble radius $R_H =
c/H_0$. We analysed the rectangular prisms with dimensions $L_x = L_y = 2
R_H$ and $L_z= 8 R_H$ (hereafter referred to as T228) and \mbox{$L_x =
L_y = 8 R_H$} and $L_z= 2 R_H$ (hereafter referred to as T882).  For
such topologies the allowed wave-vectors are of the form
\mbox{$\mathbf{k} = 2\pi (4 n_x, 4 n_y, n_z)/L_z$} and \mbox{$\mathbf{k}
= 2\pi (n_x, n_y, 4 n_z)/L_z$}, respectively, where \mbox{$n_x, n_y, n_z
\in \mathbb{Z}$}. In \sect\ref{sec:pdf} we also show some results for
the toruses with dimensions \mbox{$L_x = L_y = 4 R_H$} and $L_z= 8 R_H$
(hereafter referred to as T448) and \mbox{$L_x = L_y = 6 R_H$} and $L_z=
8 R_H$ (hereafter referred to as T668). The longer dimensions are
significantly bigger than the diameter of the observable universe (i.e.,
$6.36\ R_H$) therefore, in practice the topologies are
indistinguishable from the so-called the chimney and slab spaces,
respectively \citep{adams:2001}.

\section{Results} \label{sec:results}

We do not have an analytical expression for either the distribution of
the multipole vectors on the sky for a multi-connected universe or for
the probability distribution function (PDF) of the statistics.
Therefore we based our analysis on simulated CMB maps for each of the
studied topologies with the same cosmological parameters, and in
particular use $10^5$ -- for the distribution of the vectors on the
sky, and $10^6$ -- for the PDF of the statistics.  The signatures of a
multiconnected topology are most clearly visible for the low order
multipoles, so we confined the analysis only to the multipoles of
order $\ell \in [2,8]$. For such a range of multipoles we have 21
different pairs of multipoles for the multipole vector statistics.  We
did not add noise in the simulations, because the WMAP detector noise
is completely negligible for the low order multipoles. The observed
values of the statistics are given for the ILC \emph{WMAP} 5 years map
\citep{hinshaw:2009}.

\subsection{Distribution of the multipole vectors} \label{sec:distribution}

\begin{figure*}
\epsfig{file=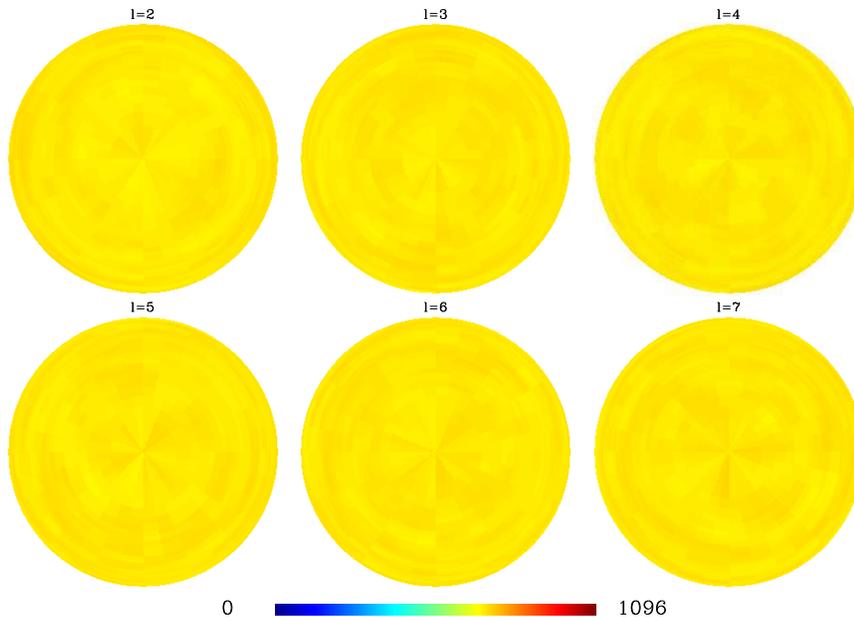,width=0.55\linewidth,angle=90}

\caption{The distribution of the multipole vectors (in orthographic
 projection) in logarithmic scale for the simply-connected universe.}
\label{fig:dist_iso}
\end{figure*}

In the case of a simply-connected universe we do not observe any specific
direction in the distribution of the CMB anisotropy.  As a consequence,
the multipole vectors (as well the cross products) are uniformly
distributed on the sky (see\footnote{because the multipole vectors do
not have their own identity, in all figures there are shown
distributions of one of the vectors for a given multipole. The
distributions of the other vectors are the same}
\fig\ref{fig:dist_iso}).

Conversely, in the case of a
multi-connected universe, if the dimension of the fundamental domain is
comparable to the universe's horizon scale, one can see preferred directions in
the distribution of the CMB anisotropy on the sky. Corresponding
multipole vectors, as well the cross products, also demonstrate non-isotropic
distributions.  The structures we observe can be qualitatively explained
as follows. For multipoles of order $\ell$, the biggest contribution
comes from the modes with the wavenumber $k \approx \ell/\eta_{\rm
LSS}$.  Significant contributions will also come from the projection on the
sphere of those modes with wavenumbers $k \gtrsim \ell/\eta_{\rm
LSS}$. Because the multipole vectors are not related linearly to the
spherical harmonic coefficients and the map of anisotropy (see appendix
\ref{sec:appendix}), it is hard to predict a priori how they are
distributed on the sky in the case of a multiconnected topology. We can,
however, expect that the distribution will somehow reflect the symmetries
of the fundamental domain.

\begin{figure*}
\epsfig{file=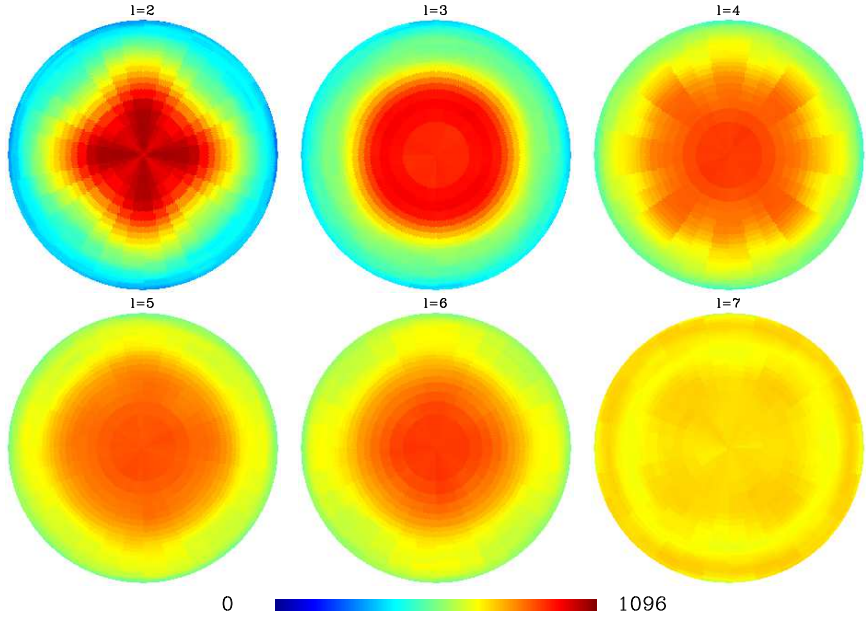,width=0.55\linewidth,angle=90}

\caption{The distribution of the multipole vectors (in orthographic
 projection) in logarithmic scale for the CMB maps of the
 multi-connected universe with the T228 topology without the ISW
 effect.}  \label{fig:dist_vec_t228}
\end{figure*}

\begin{figure*}
\epsfig{file=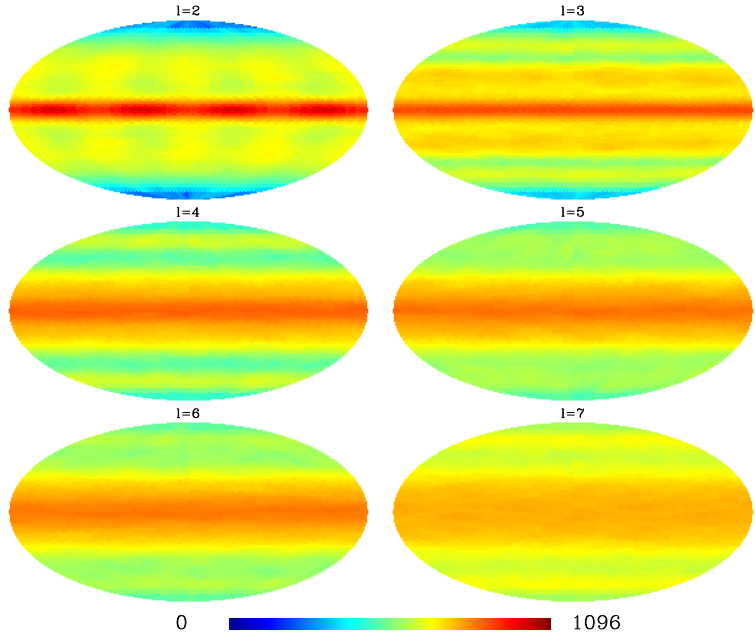,width=0.55\linewidth,angle=90}

\caption{ The distribution of the multipole vectors (in Mollweide
projection) in logarithmic scale for the CMB maps of the multi-connected
universe with the T882 topology without the ISW effect.  }
\label{fig:dist_vec_t882}
\end{figure*}

The distribution of the multipole vectors on the sky for the T228 and
T882 topologies without contribution from the ISW effect are shown in
\fig\ref{fig:dist_vec_t228} and \fig\ref{fig:dist_vec_t882},
respectively.  The distribution is related to symmetries of the
fundamental cell as expected. It is seen particularly for the T228
topology and multipoles of order $\ell=2$ and $\ell=4$.  In the case of the
quadrupole, the vectors prefer directions pointing towards those edges of
the rectangular prisms, where the power of the maps is concentrated. For
the T228 topology these are the edges parallel to the XY-plane (in the
upper and lower sides of the rectangular prism), for the T882 topology, 
the edges perpendicular to the XY-plane. Because the multipole
vectors for the quadrupole point towards the inflection points of the
CMB maps, we can deduce that the points are distributed similarly. The
distribution for higher order multipoles is less related to symmetries
of the fundamental cell. It is more azimuthally symmetric and less
concentrated in the directions preferred by the topology than is found for
the quadrupole and octopole. It is worth noticing that the volume of the
fundamental cell for the T228 topology is smaller than for the T882
topology. Thus, the non-uniform distribution of the vectors is
clearer for the former.

\begin{figure*}
\epsfig{file=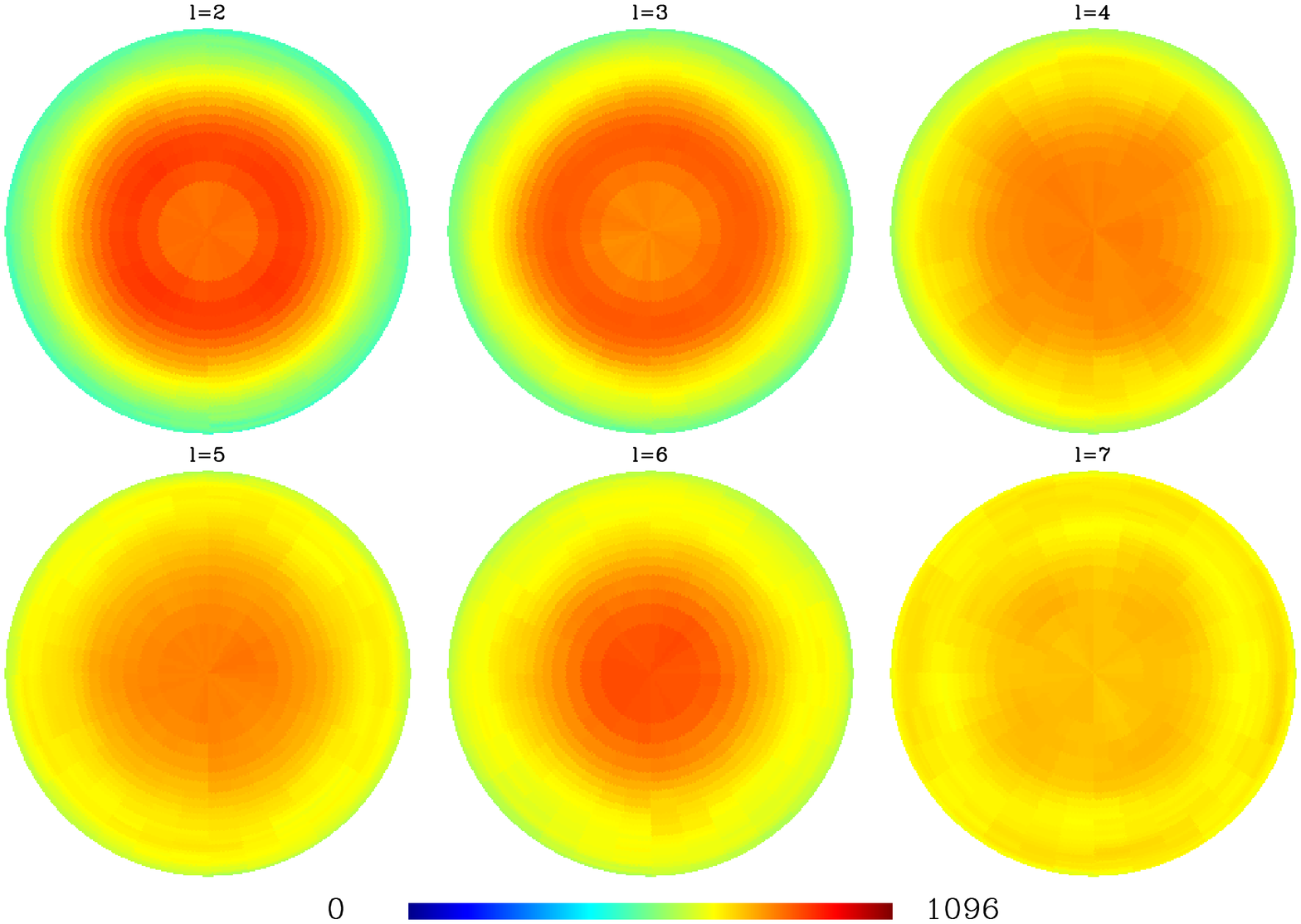,width=0.55\linewidth,angle=90}

\caption{The distribution of the multipole vectors (in orthographic
 projection) in logarithmic scale for the CMB maps of the multi-connected universe with the T228 topology 
 with the ISW effect (the cosmological constant density parameter
 $\Omega_\Lambda= 0.7$).
}
\label{fig:dist_vec_t228_isw}
\end{figure*}

\begin{figure*}
\epsfig{file=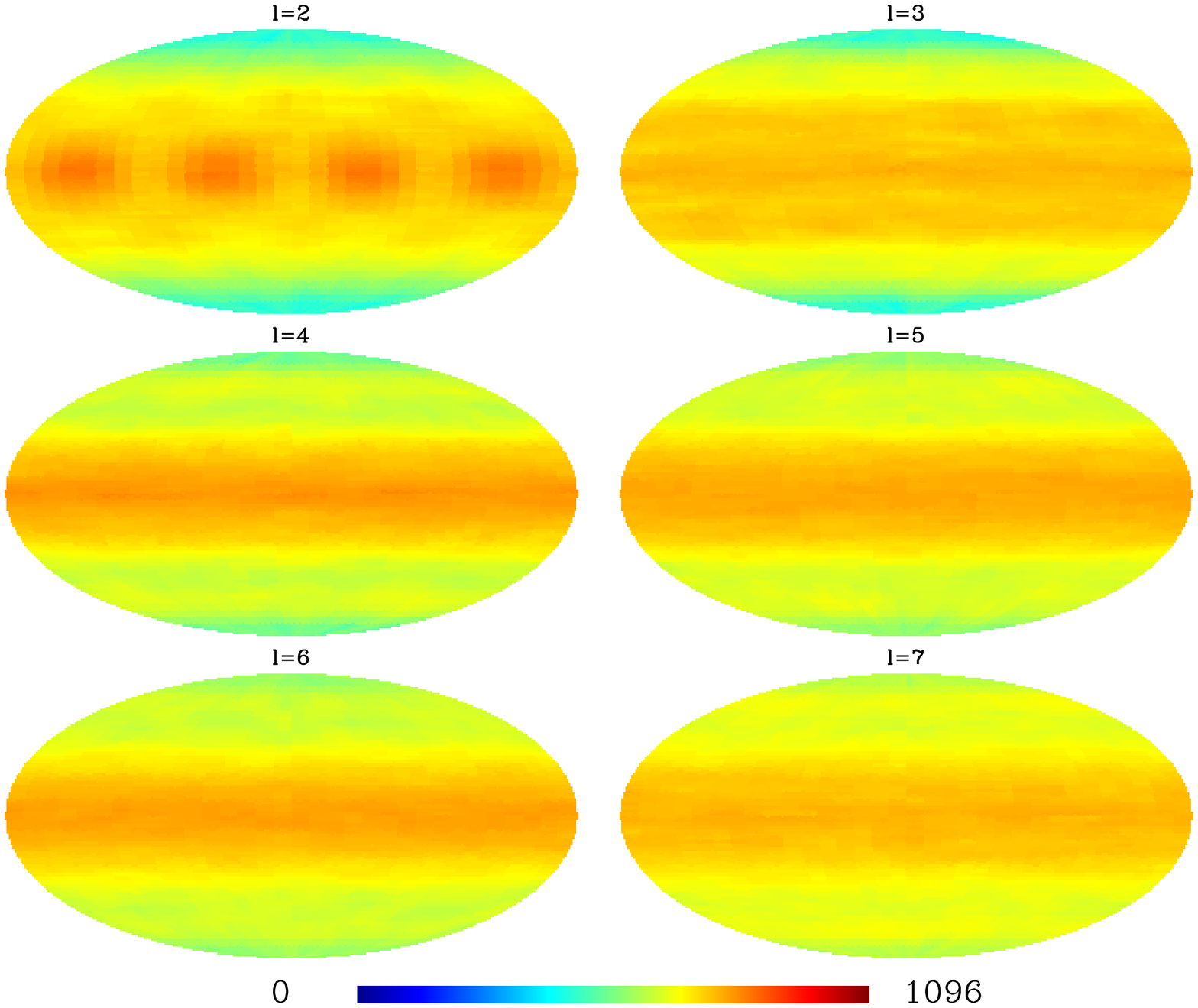,width=0.55\linewidth,angle=90}

\caption{The distribution of the multipole vectors (in Mollweide
 projection) in logarithmic scale for the CMB maps of the
 multi-connected universe with the T882 topology with the ISW effect
 (cosmological constant density parameter $\Omega_\Lambda= 0.7$).  }
 \label{fig:dist_vec_t882_isw}
\end{figure*}

The signatures of the multiconnected topology will be diluted by the
integrated Sachs-Wolfe (ISW) effect. This gives a contribution to the large
angular scales. However since it arises from the evolution of structures close
to the observer, it does not contain information about the global properties
of the universe. The influence of the ISW effect on the distribution of the
multipole vectors is seen in \fig\ref{fig:dist_vec_t228_isw} and
\fig\ref{fig:dist_vec_t882_isw}.  In both cases, the characteristic
pattern in the distribution of vectors is less pronounced than for
maps without the ISW effect.

\subsection{The probability distribution function} \label{sec:pdf}

\begin{figure*}
\epsfig{file=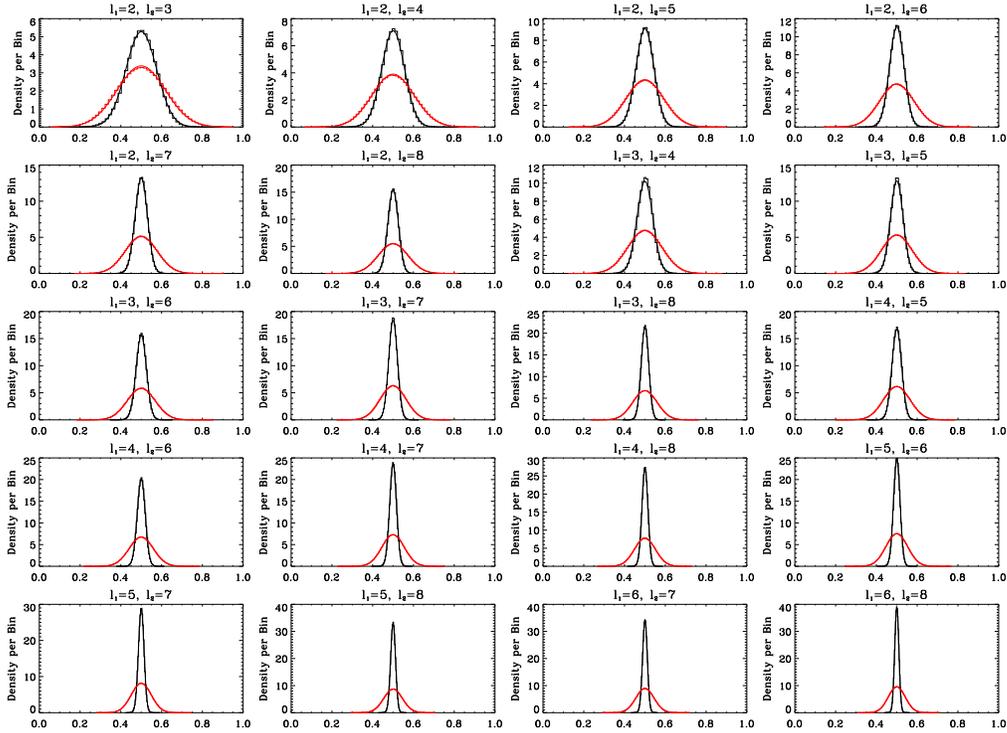,width=0.55\linewidth,angle=90}

\caption{Distribution of the multipole vectors statistic for the
simply-connected universe (black line) and the distribution of the sum
of $M$ (number of dot products in the corresponding statistic)
independent random variables (red line). There are also shown fitted to
the histograms normal distributions (solid black and red lines,
respectively).}  \label{fig:vec_iso}
\end{figure*}

The dot products of the vectors for two different multipoles are
uniformly distributed. However, the distribution of their sum does not
correspond to the distribution of the sum of $M$ independent uniformly
distributed random variables, because the multipole vectors
corresponding to the same multipole are not statistically independent
\citep{land:2005}. This is seen in \fig\ref{fig:vec_iso}. The internal
correlations of the multipole vectors cause the PDF of the
statistic to be narrower than the distribution of the sum of independent
variables. Moreover, although at first glance the distributions appear
Gaussian, they are not so.  The kurtosis of the distributions is
positive, while the skewness is slightly (though significantly) negative
for all pairs of multipoles. Deviation from Gaussianity is even stronger
in the case of universes with a multiconnected topology.

Because the PDF of the multipole vector statistics for the
multiconnected universes is similar to that for the simply-connected
universe for higher order multipoles, for the T228 and
T882 topologies we show only the distributions for those lower order multipoles with the
biggest deviation from the PDF of the simply-connected universe.

\begin{figure*}
\epsfig{file=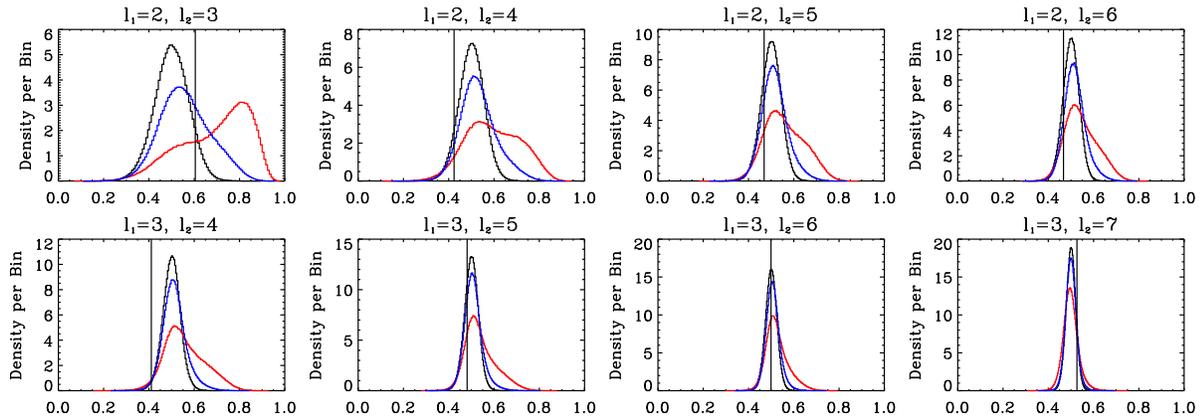,width=0.9\linewidth}

\caption{Distribution of the $S_{\rm vv}$ statistic for the
 simply-connected (black lines) and the multi-connected universe with
 the T228 topology without (red line) and with the ISW effect (blue
 line) (the cosmological constant density parameter $\Omega_\Lambda
 =0.7$).  Vertical line indicates value of the statistic for the ILC
 \emph{WMAP} 5 years map.}  \label{fig:vec_vec_t228_t228isw_iso}
\end{figure*}

\begin{figure*}
\epsfig{file=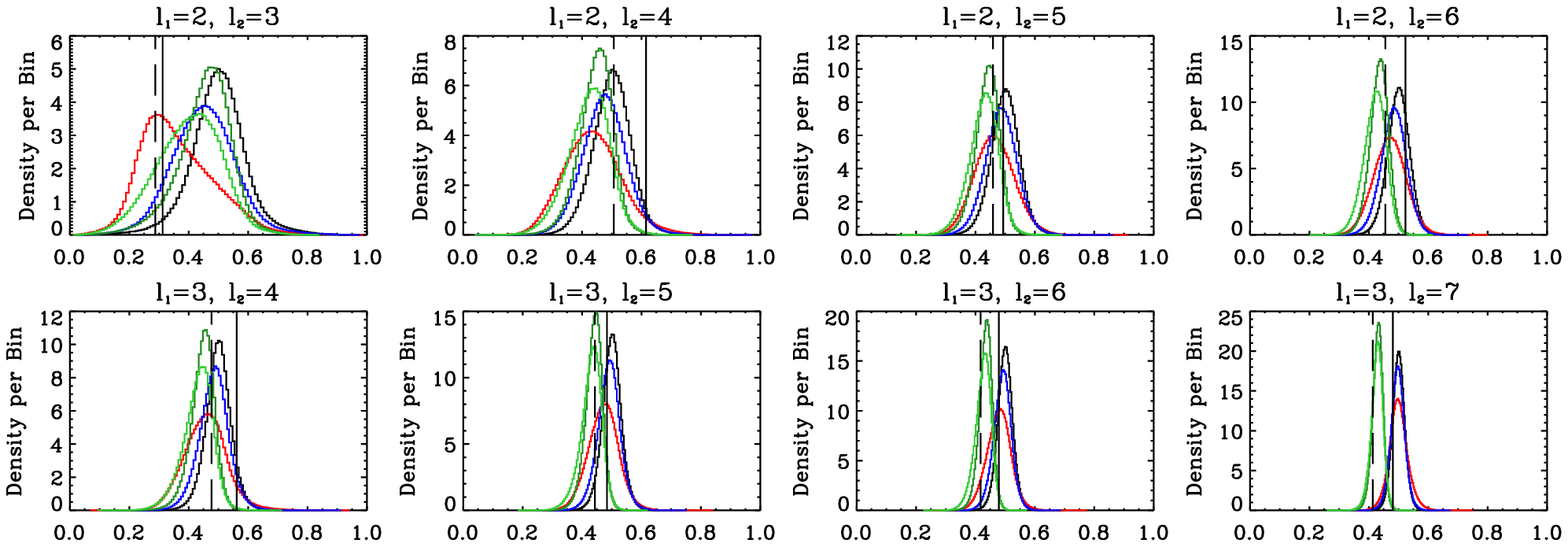,width=0.9\linewidth}

\caption{Distribution of the $S_{\rm vw}$ and $S_{\rm vw}^u$ statistics
 for the simply-connected (black and dark green lines, respectively) and
 multi-connected universe with the T228 topology without (red line, only
 for the $S_{\rm vw}$ statistic) and with the ISW effect (blue and light
 green lines, respectively) (the cosmological constant density parameter
 $\Omega_\Lambda =0.7$).
Vertical lines indicate values of the statistics for the ILC \emph{WMAP} 5
 years map: solid line for $S_{\rm vw}$ and dashed line for $S_{\rm vw}^u$
 statistic.}
\label{fig:vec_cross_t228_t228isw_iso}
\end{figure*}

\begin{figure*}
\epsfig{file=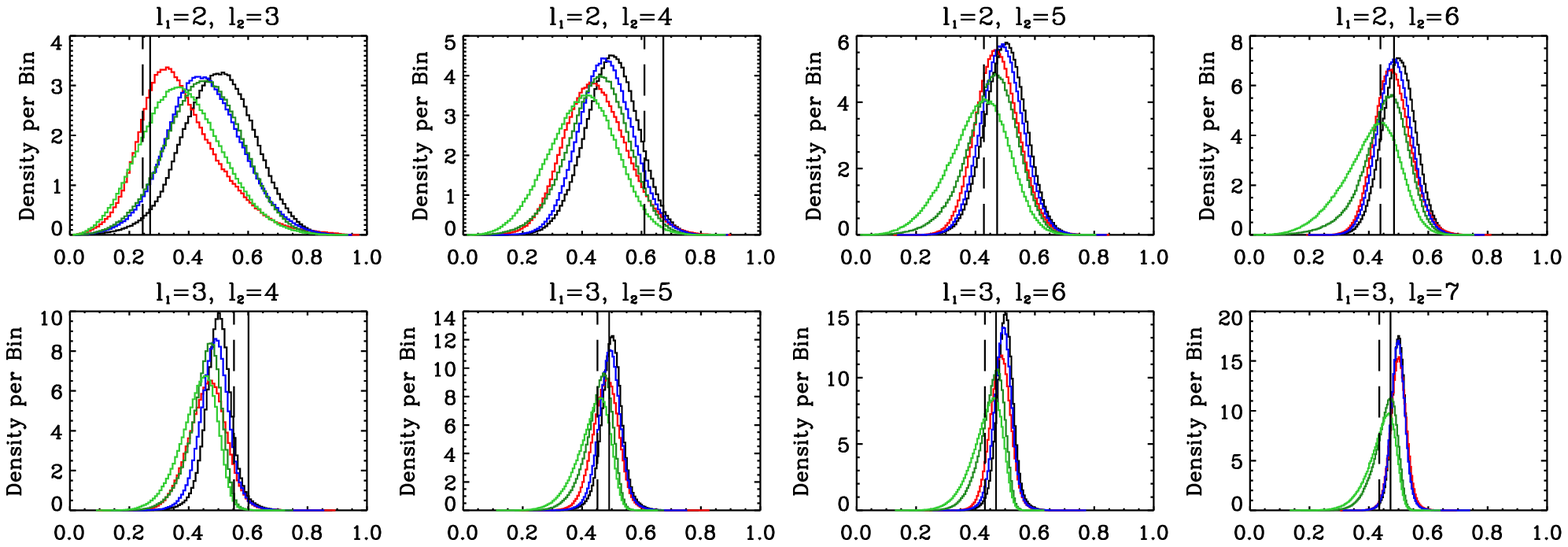,width=0.9\linewidth}

\caption{Distribution of the $S_{\rm wv}$ and $S_{\rm wv}^u$ statistics
 for the simply-connected (black and dark green lines, respectively) and
 multi-connected universe with the T228 topology without (red line, only
 for the $S_{\rm wv}$ statistic) and with the ISW effect (blue and light
 green lines, respectively) (the cosmological constant density parameter
 $\Omega_\Lambda =0.7$).  Vertical lines indicate values of the
 statistics for the ILC \emph{WMAP} 5 years map: solid line for $S_{\rm
 wv}$ and dashed line for $S_{\rm wv}^u$ statistic.}
 \label{fig:cross_vec_t228_t228isw_iso}
\end{figure*}

\begin{figure*}
\epsfig{file=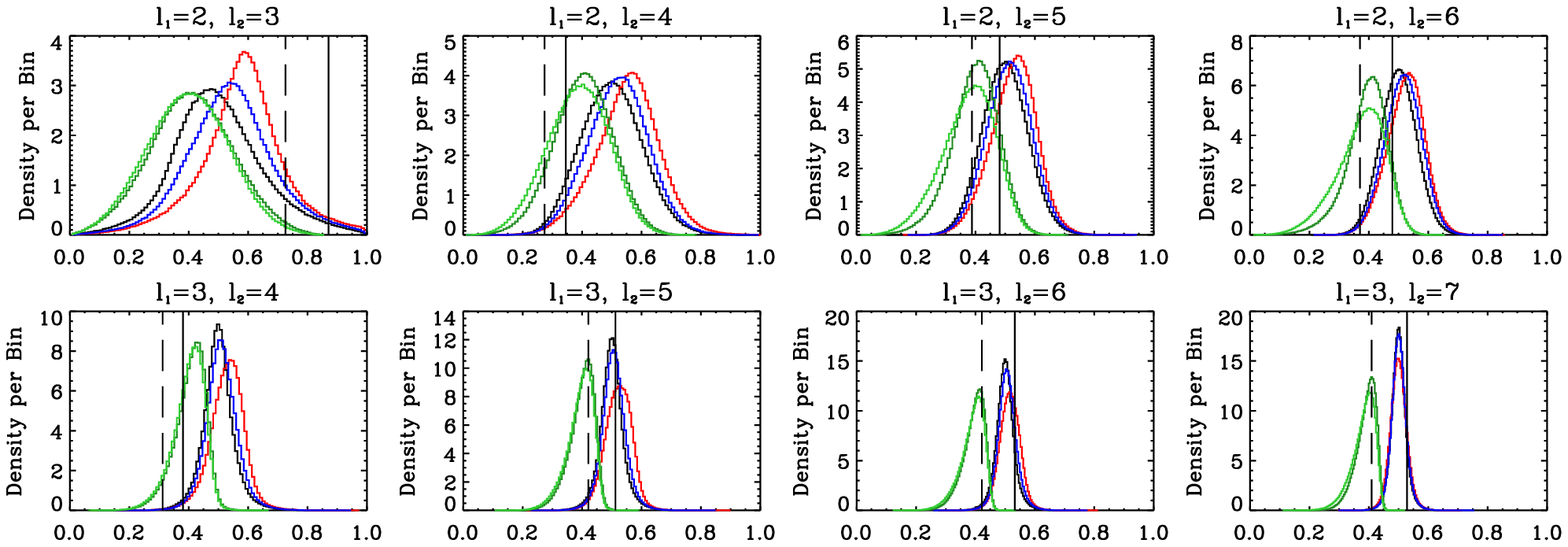,width=0.9\linewidth}

\caption{Distribution of the $S_{\rm ww}$ and $S_{\rm ww}^u$ statistics
 for the simply-connected (black and dark green lines, respectively) and
 multi-connected universe with the T228 topology without (red line, only
 for the $S_{\rm ww}$ statistic) and with the ISW effect (blue line and
 light green lines, respectively) (the cosmological constant density
 parameter $\Omega_\Lambda =0.7$).  Vertical lines indicate value of the
 statistics for the ILC \emph{WMAP} 5 years map: solid line for $S_{\rm
 ww}$ and dashed line for $S_{\rm ww}^u$ statistic.}
 \label{fig:cross_cross_t228_t228isw_iso}
\end{figure*}

For the T228 topology the alignment of the multipole vectors in the direction of
the longer side of the prism causes a significant
shift of the PDF of the $S_{\rm vv}$ statistic toward higher values (see
\fig\ref{fig:vec_vec_t228_t228isw_iso}). Because cross products of the
multipole vectors are orthogonal to the vectors, alignment of the
multipole vectors results in the distributions of the $S_{\rm vw}$ and
$S_{\rm wv}$ statistics to be shifted toward zero (see
\fig\ref{fig:vec_cross_t228_t228isw_iso} and
\fig\ref{fig:cross_vec_t228_t228isw_iso}). On the other hand, alignment
of the cross products in the direction of the shorter sides of the prism and in
the plane perpendicular to the longer side means that the distribution of the 
$S_{\rm ww}$ statistic is slightly moved toward one (see
\fig\ref{fig:cross_cross_t228_t228isw_iso}).

\begin{figure*}
\epsfig{file=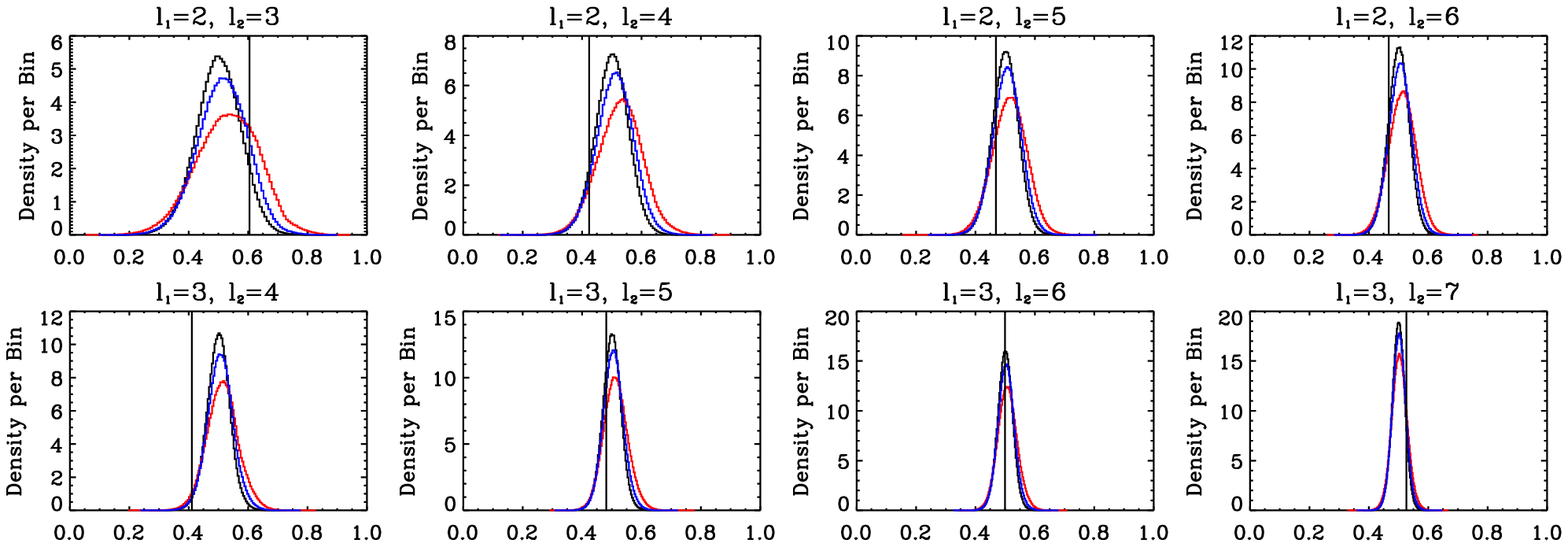,width=0.9\linewidth}

\caption{Distribution of the $S_{\rm vv}$ statistic for the
 simply-connected (black lines) and multi-connected universe with
 topology $T882$ without the ISW effect (red line) and with the ISW
 effect (the cosmological constant density parameter $\Omega_\Lambda
 =0.7$) (blue line). Vertical line indicates value of the statistic for
 the ILC \emph{WMAP} 5 years map.}  \label{fig:vec_vec_t882_t882isw_iso}
\end{figure*}

\begin{figure*}
\epsfig{file=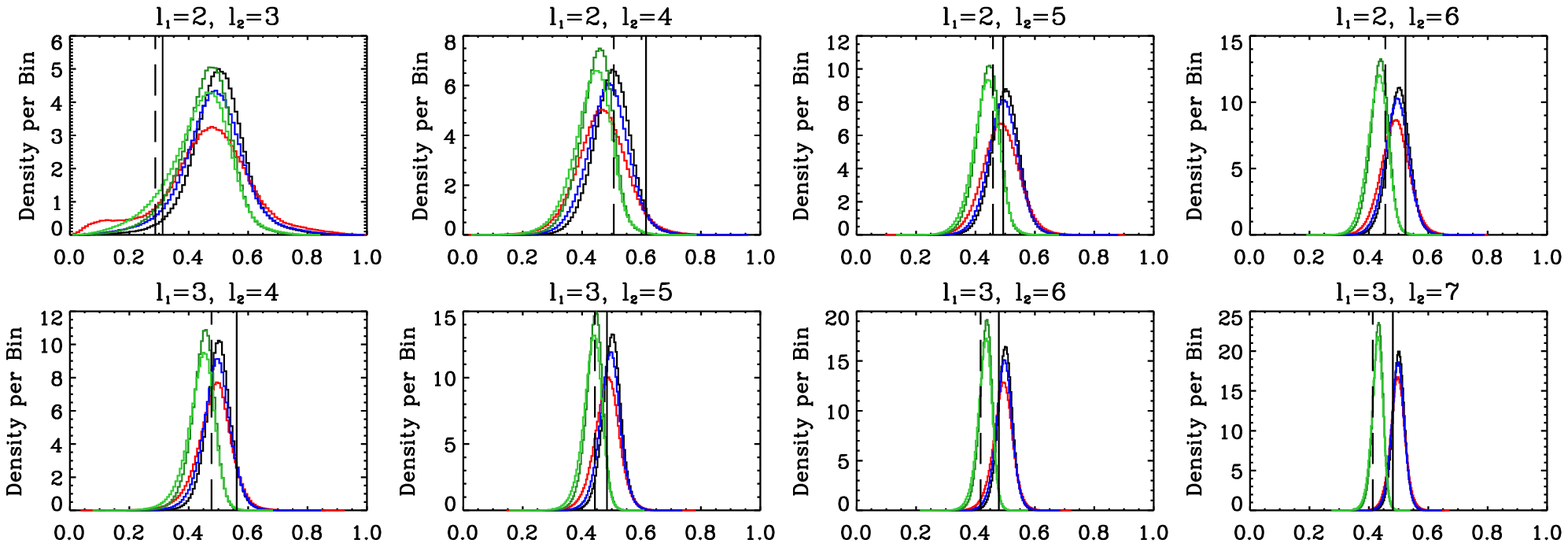,width=0.9\linewidth}

\caption{Distribution of the $S_{\rm vw}$ and $S_{\rm vw}^u$ statistics
 for the simply-connected (black and dark green lines, respectively) and
 multi-connected universe with the T882 topology without the ISW effect
 (red line, only for the $S_{\rm vw}$ statistic) and with the ISW effect
 (blue and light green lines, respectively) (the cosmological constant
 density parameter $\Omega_\Lambda =0.7$) . Vertical lines indicate
 values of the statistics for the ILC \emph{WMAP} 5 years map: solid
 line for $S_{\rm vw}$ and dashed line for $S_{\rm vw}^u$ statistic.}
 \label{fig:vec_cross_t882_t882isw_iso}
\end{figure*}

\begin{figure*}
\epsfig{file=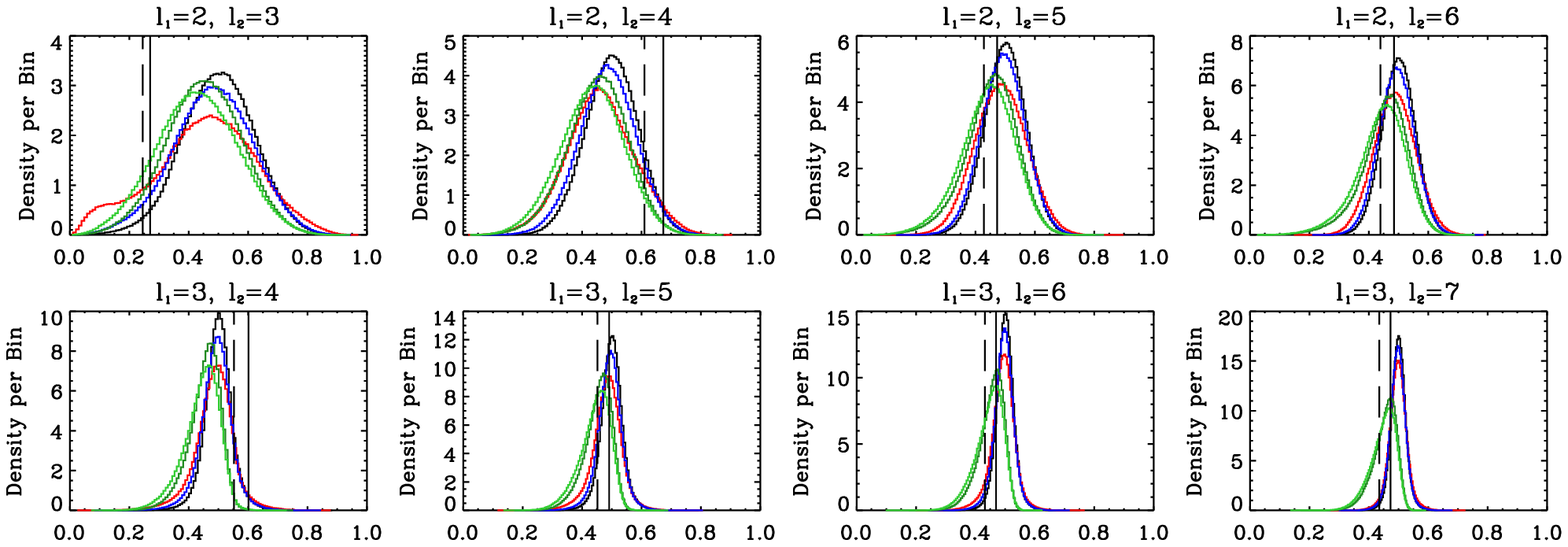,width=0.9\linewidth}

\caption{Distribution of the $S_{\rm wv}$ and $S_{\rm wv}^u$ statistics
 for the simply-connected (black and dark green lines, respectively) and
 multi-connected universe with the T882 topology without the ISW effect
 (red line, only for the $S_{\rm wv}$ statistic) and with the ISW effect
 (blue and light green lines, respectively) (the cosmological constant
 density parameter $\Omega_\Lambda =0.7$). Vertical lines indicate
 values of the statistics for the ILC \emph{WMAP} 5 years map: solid
 line for $S_{\rm wv}$ and dashed line for $S_{\rm wv}^u$ statistic.}
 \label{fig:cross_vec_t882_t882isw_iso}
\end{figure*}

\begin{figure*}
\epsfig{file=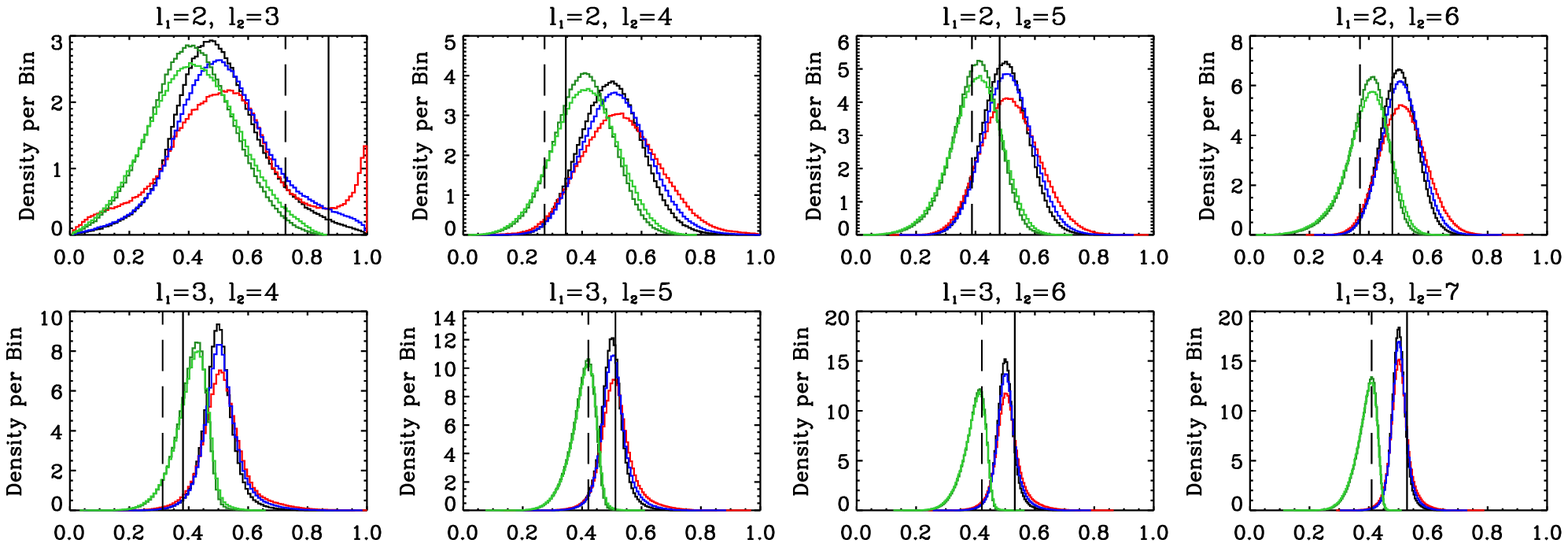,width=0.9\linewidth}

\caption{Distribution of the $S_{\rm ww}$ and $S_{\rm ww}$ statistics
 for the simply-connected (black and dark green lines, respectively) and
 multi-connected universe with the T882 topology without the ISW effect
 (red line, only for the $S_{\rm ww}$ statistic) and with the ISW effect
 (blue and light green lines, respectively) (the cosmological constant
 density parameter $\Omega_\Lambda =0.7$) Vertical lines indicate values
 of the statistics for the ILC \emph{WMAP} 5 years map: solid line for
 $S_{\rm ww}$ and dashed line for $S_{\rm ww}^u$ statistic.}
 \label{fig:cross_cross_t882_t882isw_iso}
\end{figure*}

Similar tendencies in the PDF are seen also in case of the T882
topology (see \fig\ref{fig:vec_vec_t882_t882isw_iso},
\ref{fig:vec_cross_t882_t882isw_iso}, \ref{fig:cross_vec_t882_t882isw_iso}
and \ref{fig:cross_cross_t882_t882isw_iso}). It can be explained by the alignment of the multipole vectors in
the plane spanned by the longer sides and the cross products in the
direction of the shorter side. However, because of the larger volume of the
fundamental domain than for the T228 topology and weaker alignment of
the vectors, the signatures of the topology are less significant. Only
in the case of the $S_{\rm vw}$, $S_{\rm wv}$ and $S_{\rm ww}$ statistics,
for the pair of multipoles of order $\ell=2$ and $\ell=3$ (without
contribution from the ISW effect), does the distribution demonstrate a bigger
deviation from the PDF of the simply-connected universe. It has a
pronounced local maximum around one for the $S_{\rm ww}$ statistic (see
\fig\ref{fig:cross_cross_t882_t882isw_iso}) and
show some excess close to zero for the $S_{\rm vw}$ and $S_{\rm wv}$
statistics (see \fig\ref{fig:vec_cross_t882_t882isw_iso} and \ref{fig:cross_vec_t882_t882isw_iso}).

The statistics corresponding to the unnormalised cross products take
smaller values in comparison to those with normalised ones, because
of the smaller norm of the unnormalised cross products. It is seen
especially well for higher order multipoles. 

As in the case of the distribution of the vectors on the sky, the ISW effect
significantly dilutes the signatures of the topology for all kind of
statistics.

For a few pairs of multipoles, especially for $(2,3)$, $(2,4)$, $(3,4)$
and $(3,7)$, the observed values of the statistics differ from the
predictions for the simply-connected universe substantially. However,
not all of them indicate a multiconnected topology. While correlations
for the pairs $(2,3)$ and $(3,7)$ prefer one of the topologies under consideration,
correlations for the pairs $(2,4)$ and $(3,4)$ are in conflict with
these models. A quantitative estimation of the goodness of fit of the models
to the data is presented in \sect\ref{sec:probabilities}.

\begin{figure*}
\begin{tabular}{cc}
 \epsfig{file=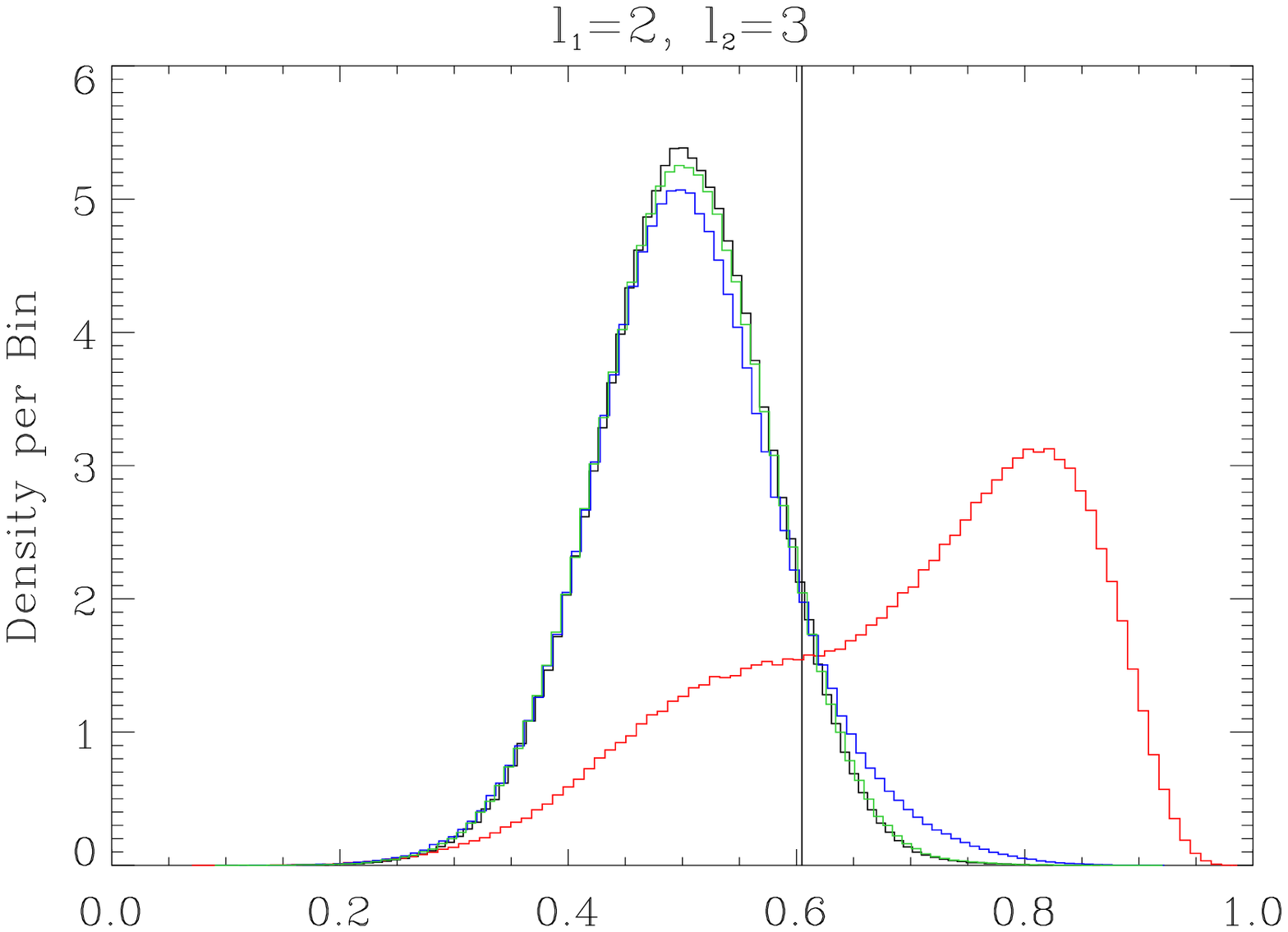,width=0.4\linewidth} &
 \epsfig{file=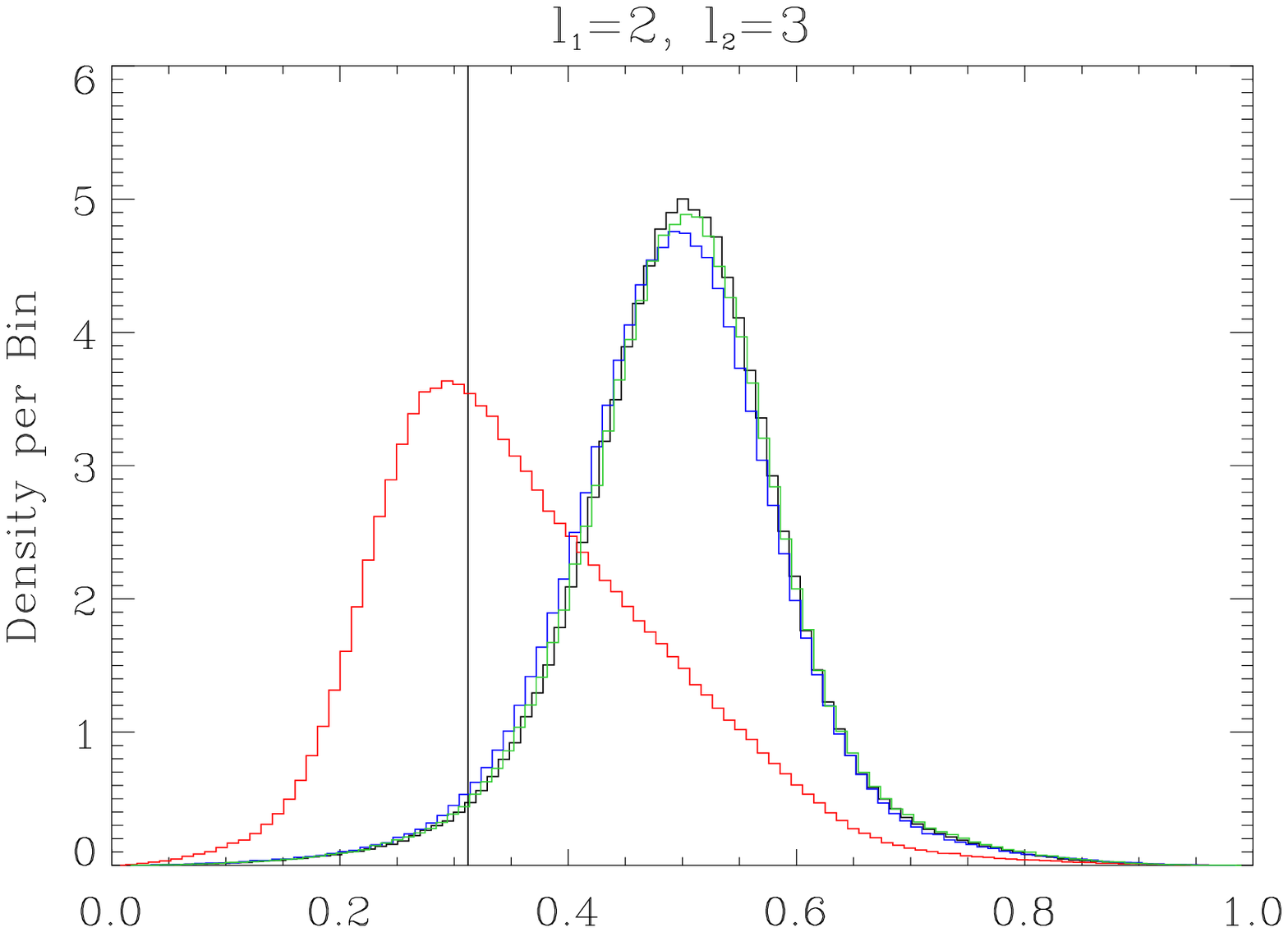,width=0.4\linewidth} \\
 \epsfig{file=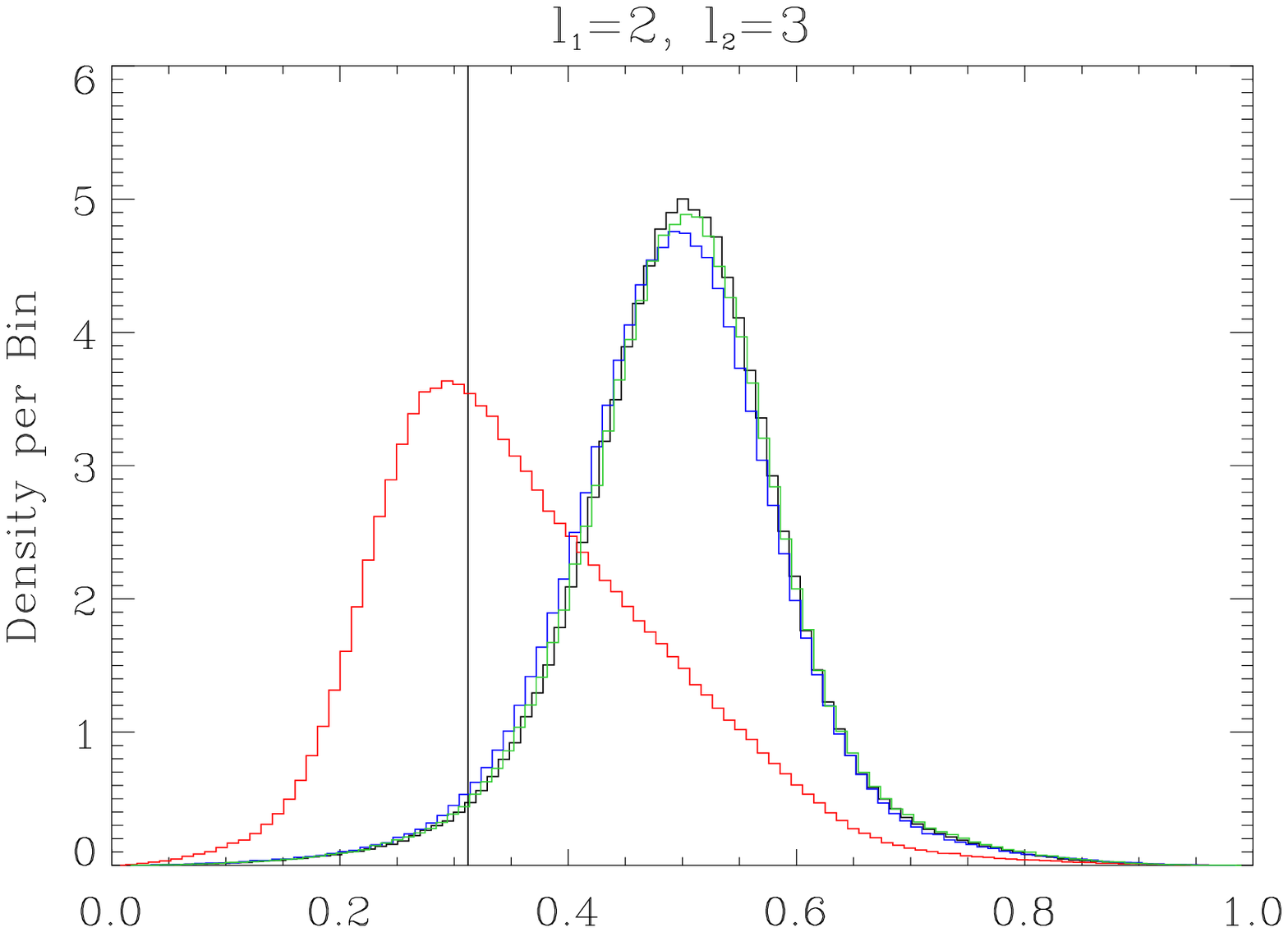,width=0.4\linewidth} &
 \epsfig{file=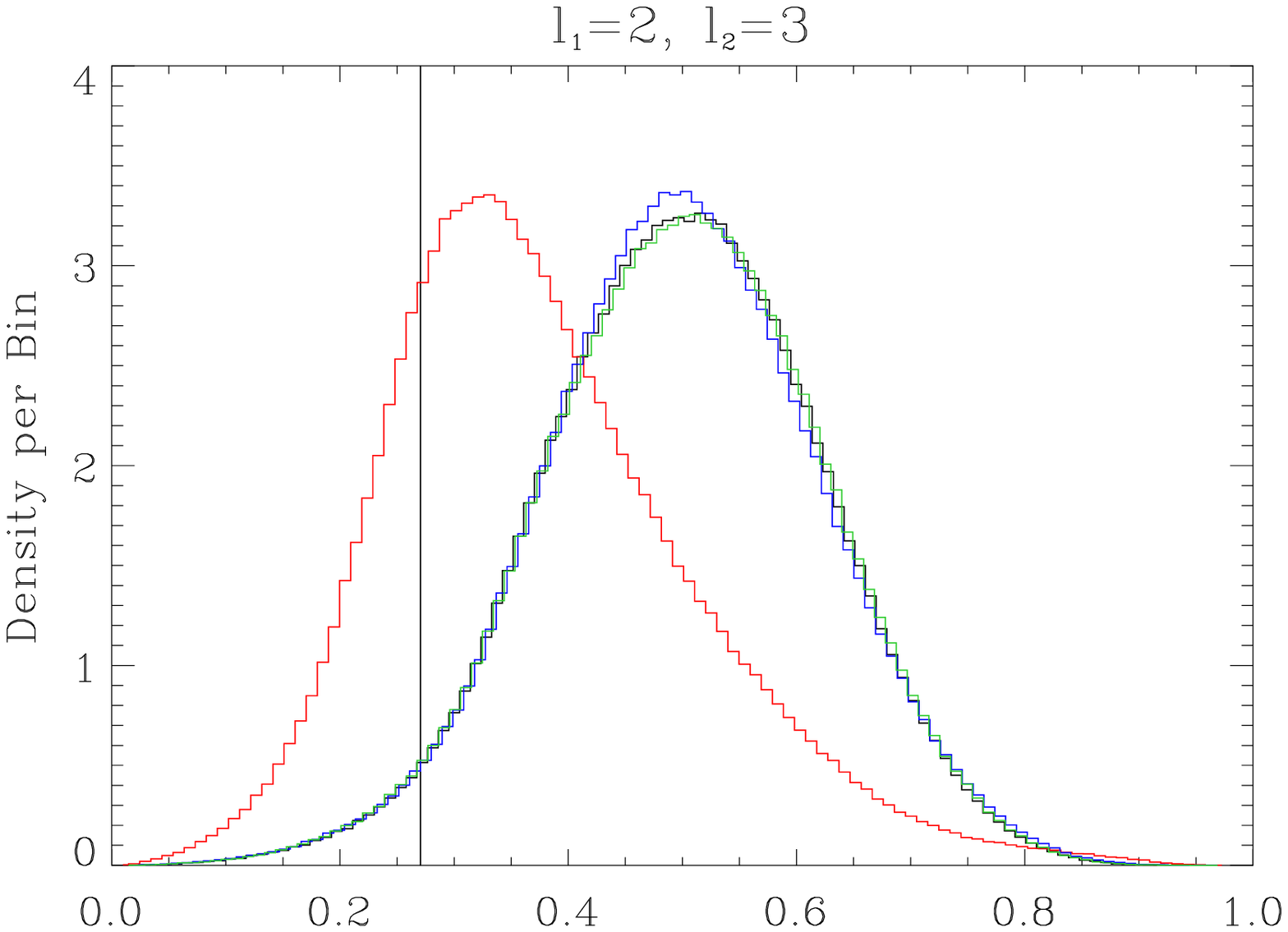,width=0.4\linewidth} \\
\end{tabular}

\caption{ Distribution for pairs of multipoles $(\ell_1,\ell_2) = (2,3)$
of the $S_{\rm vv}$, $S_{\rm ww}$ (the left and right figure in the
upper row, respectively), $S_{\rm vw}$ and $S_{\rm wv}$ (the left and
right figure in the lower row, respectively) statistics for the
simply-connected (black lines) and multi-connected universe with the
T228 (red line), T448 (blue line) and T668 (green line) topology without
the ISW effect. Vertical line indicates value of the statistic for the
ILC \emph{WMAP} 5 years map.  } \label{fig:dist_txx8_sw}
\end{figure*}

To study the dependence of the PDF of the statistics on the dimensions of the
fundamental domain, we estimated the PDF also for the 3-toruses T448 and
T668. The distribution of the statistics for the pair of multipoles
$(\ell_1, \ell_2) = (2,3)$ without any contribution from the ISW effect is
shown in \fig\ref{fig:dist_txx8_sw}. It can be seen that traces of a
multiconnected topology are substantially degraded for a universe with
larger dimensions, and almost completely disappears already for the
topology T448. For the simulations including the ISW effect, the
statistics are even less sensitive to the dimensions of the fundamental cell.

\subsection{Probabilities} \label{sec:probabilities}

\begin{figure*}
\begin{tabular}{cc}
 \epsfig{file=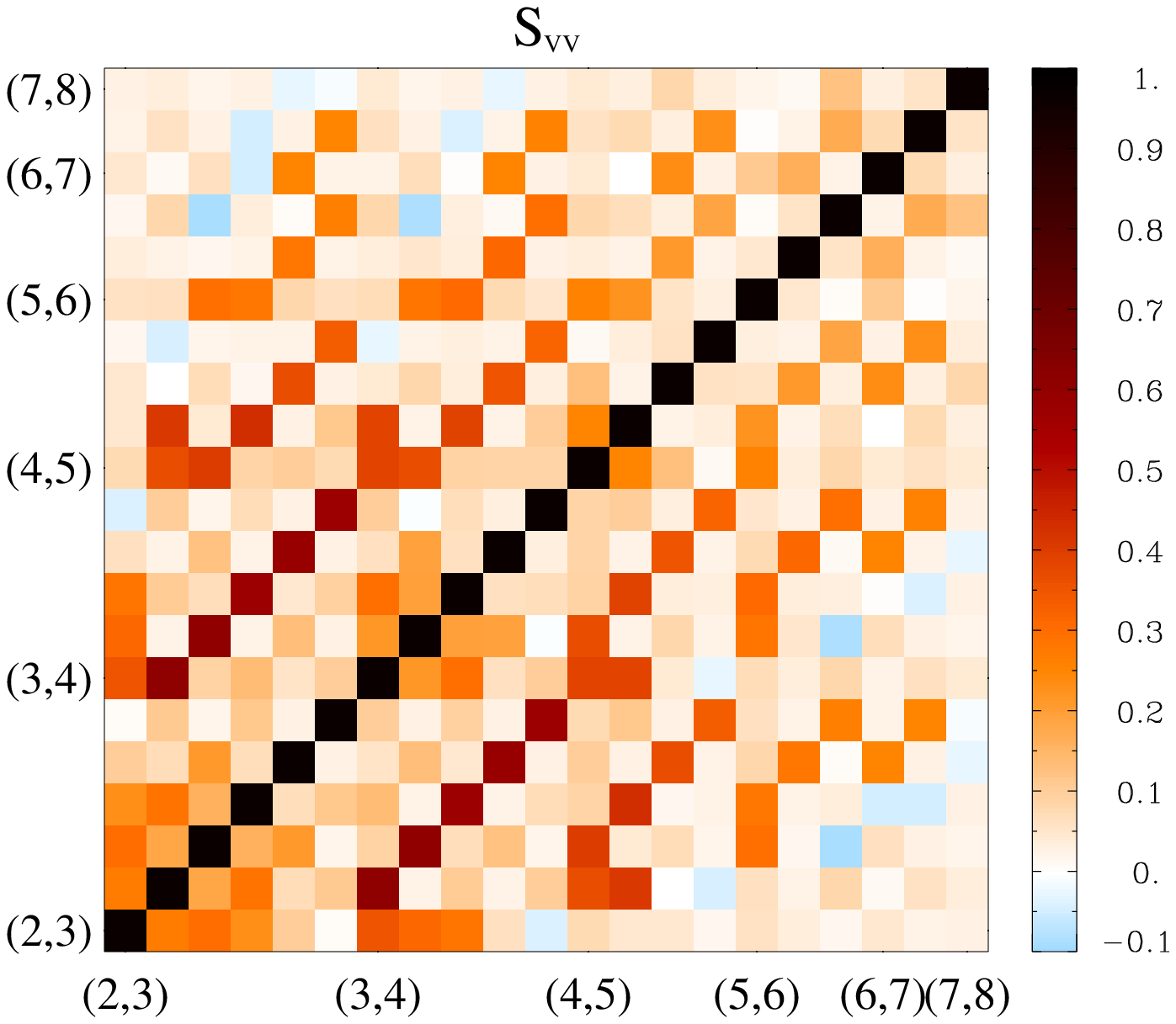,width=0.4\linewidth} &
 \epsfig{file=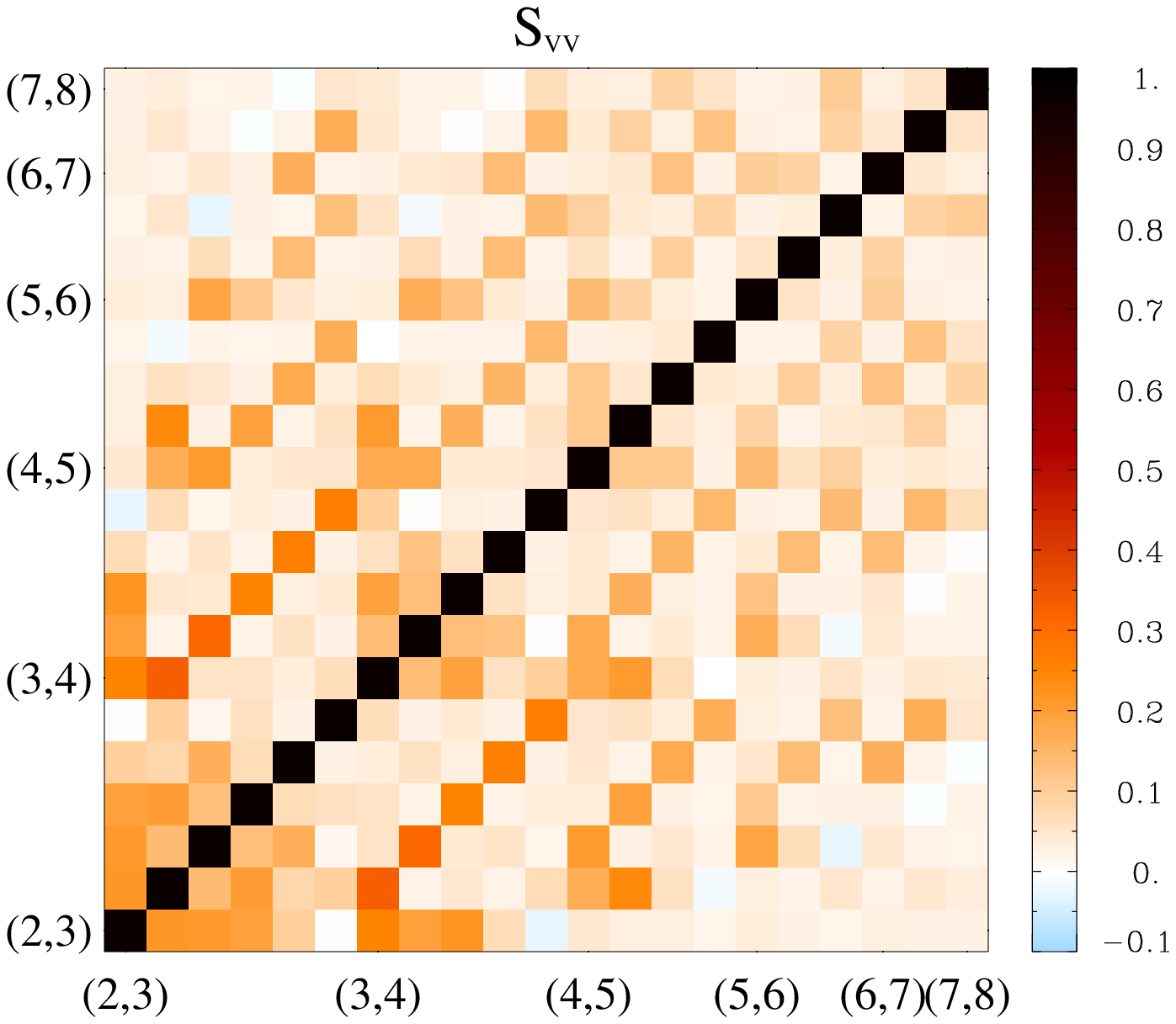,width=0.4\linewidth} \\
 \epsfig{file=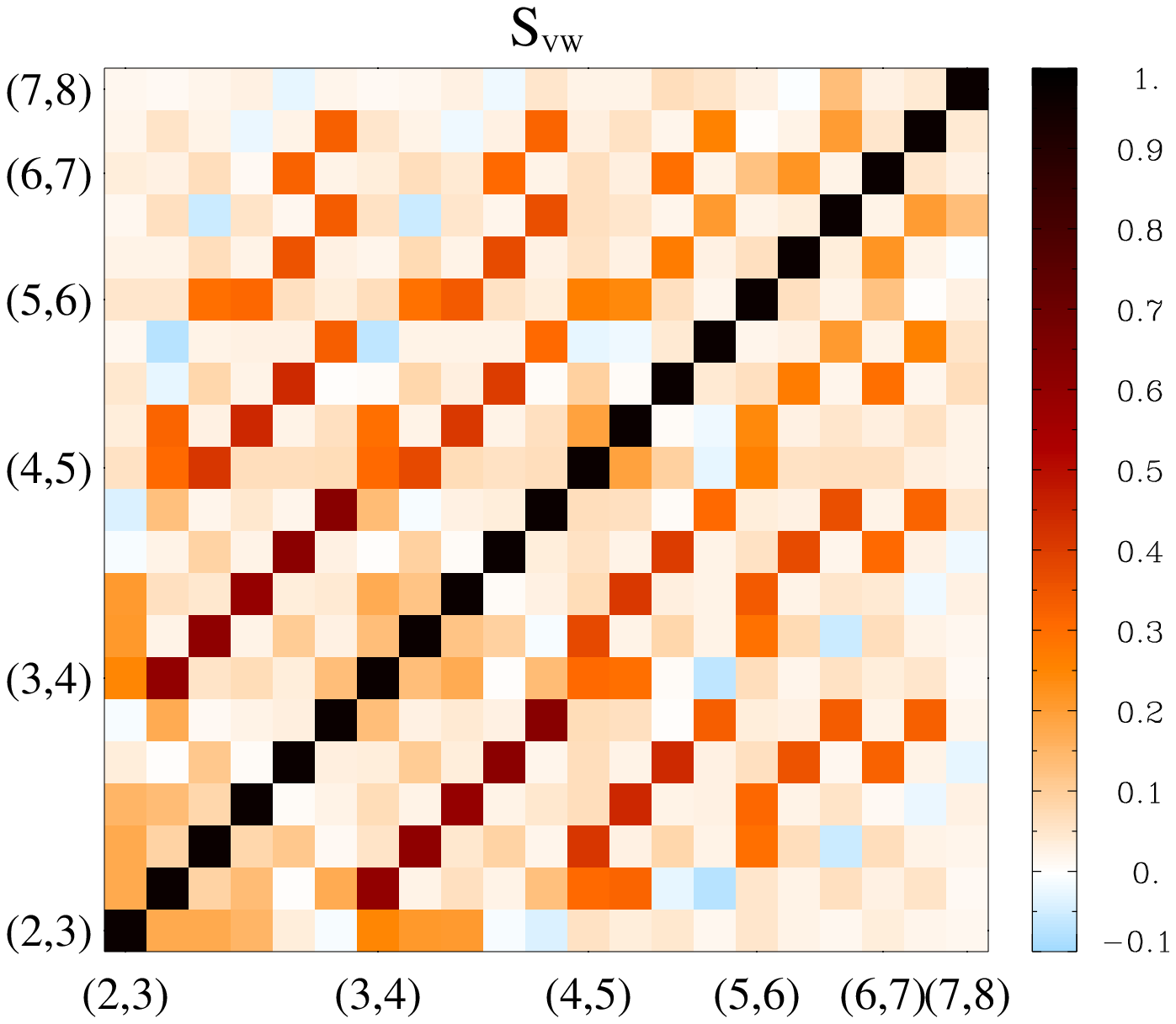,width=0.4\linewidth} &
 \epsfig{file=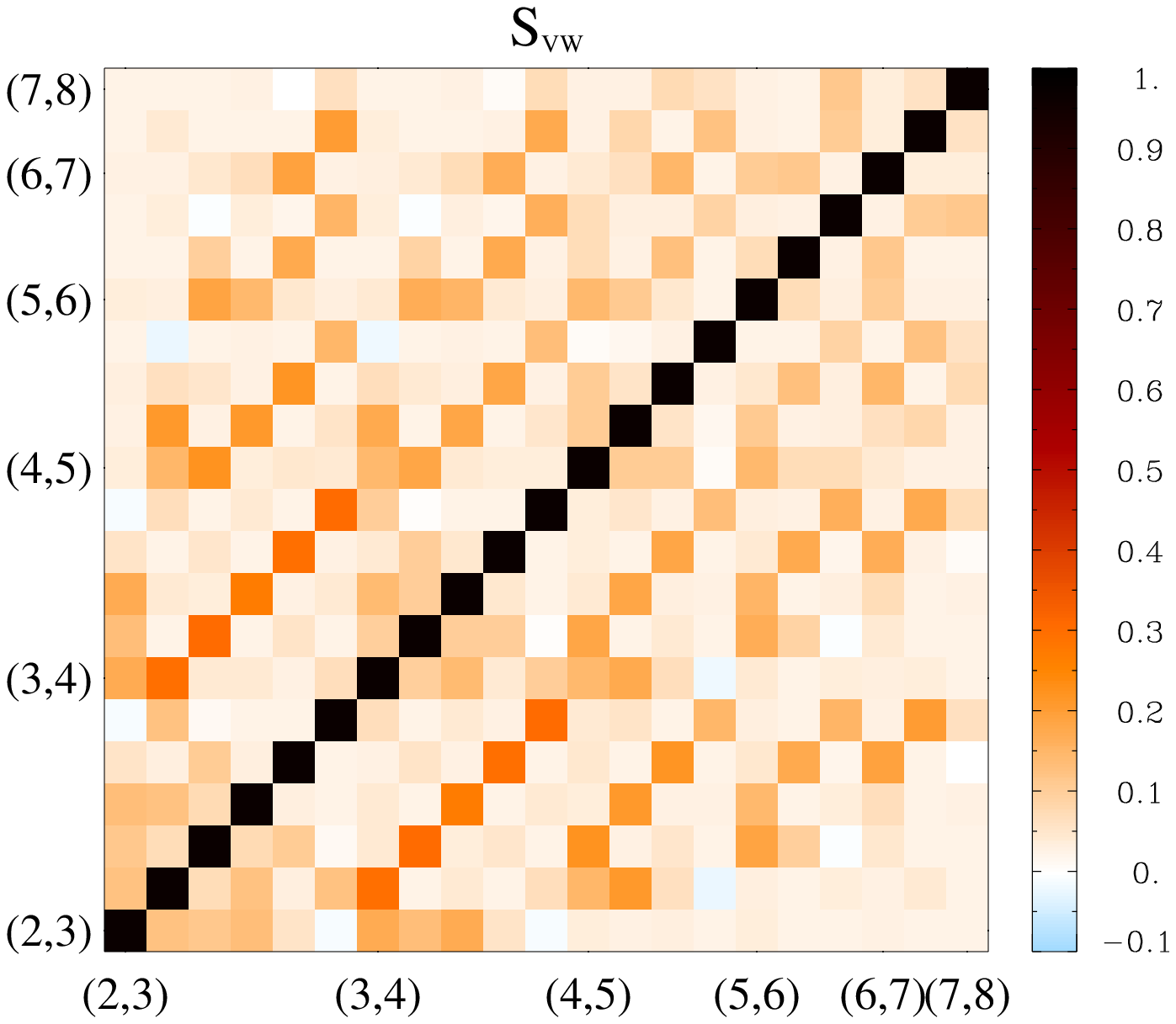,width=0.4\linewidth} \\
 \epsfig{file=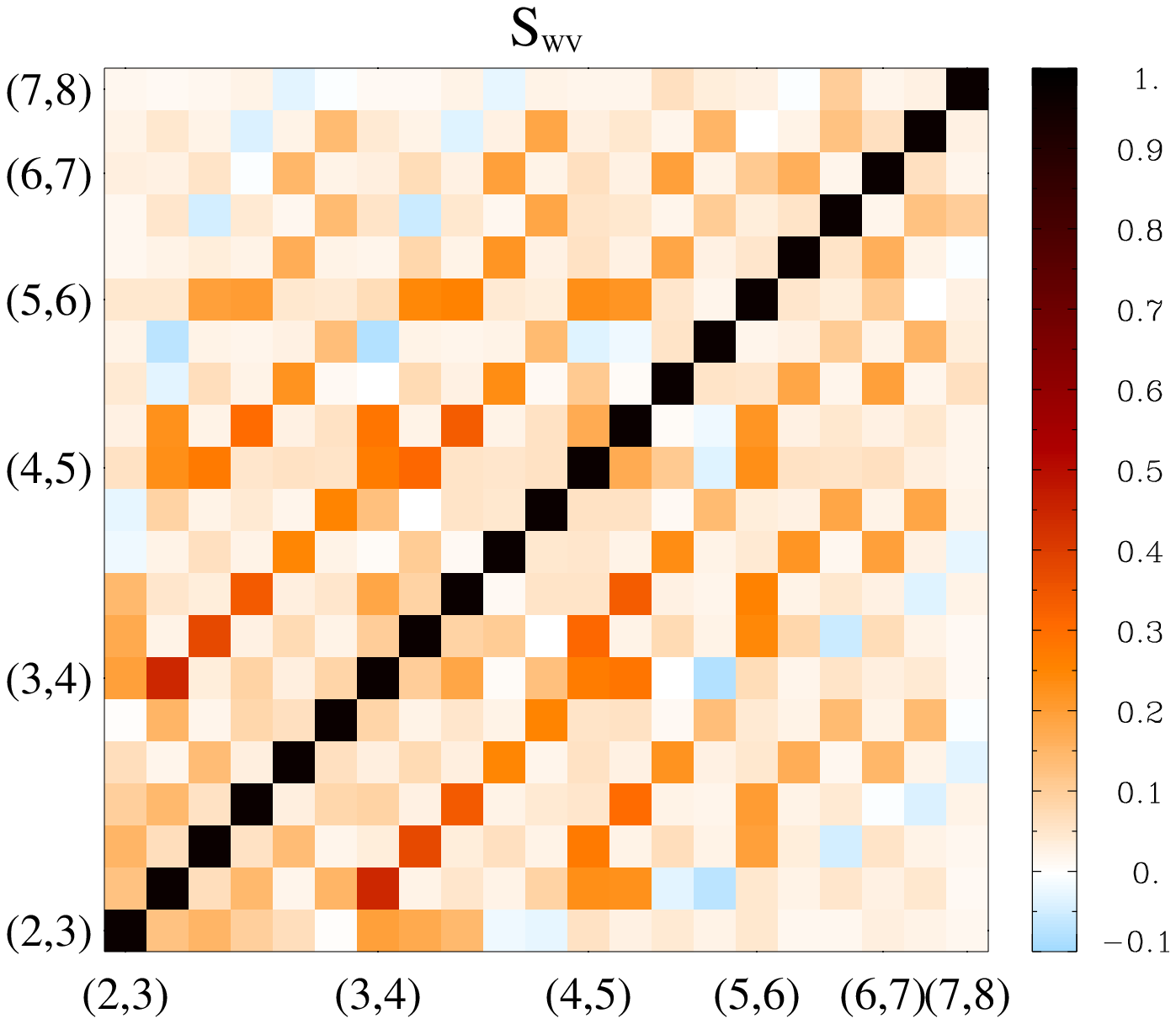,width=0.4\linewidth} &
 \epsfig{file=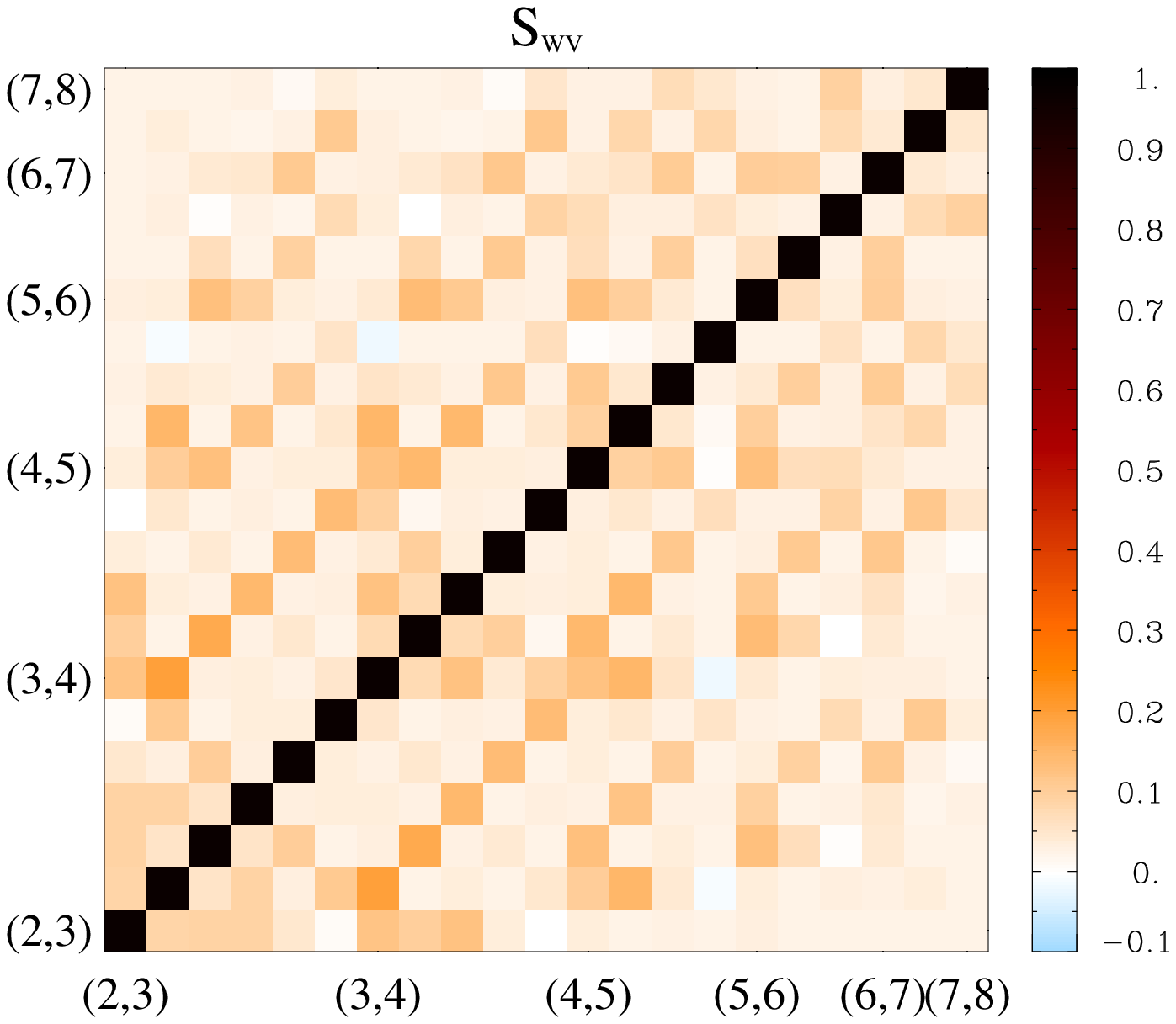,width=0.4\linewidth} \\
 \epsfig{file=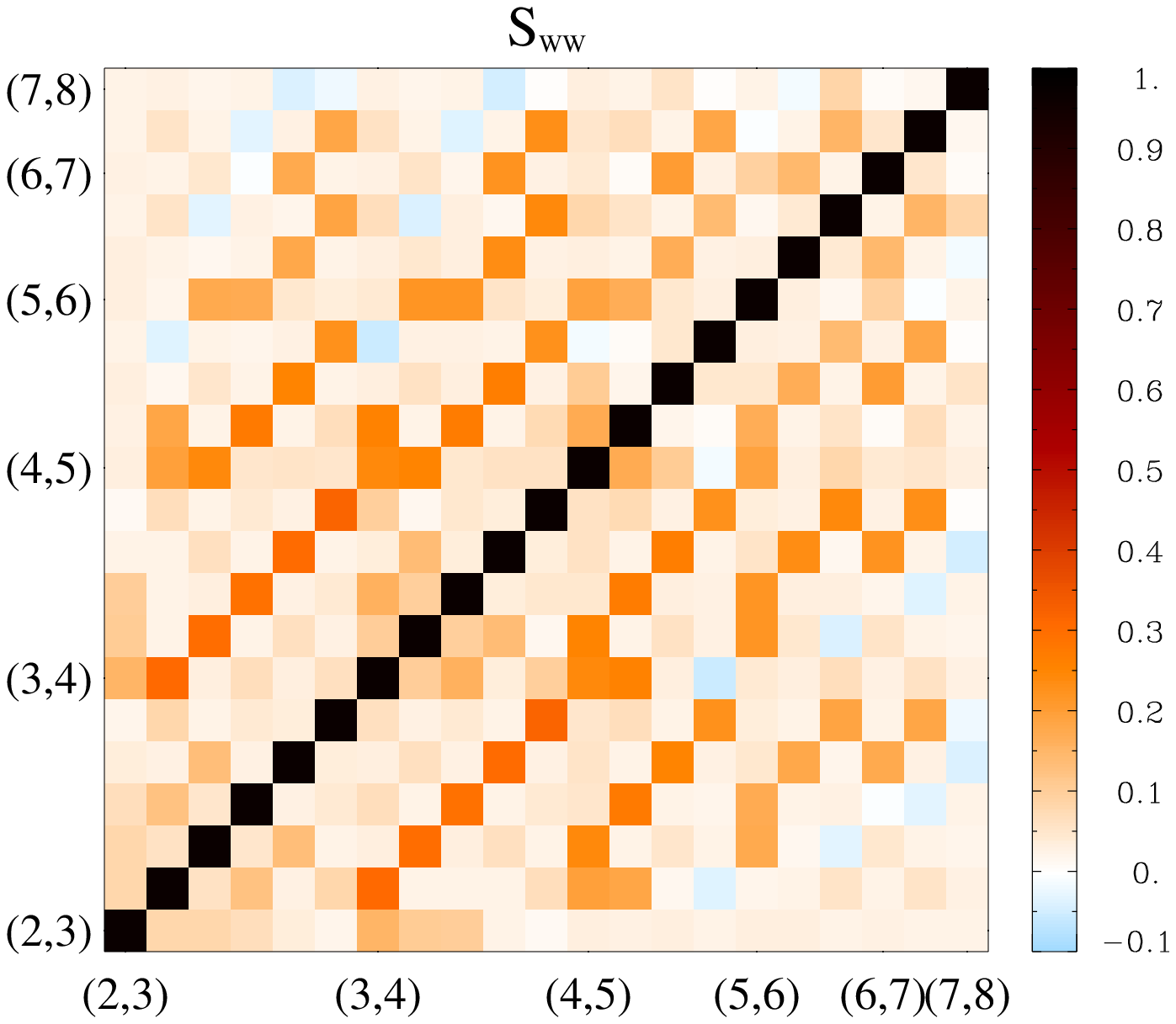,width=0.4\linewidth} &
 \epsfig{file=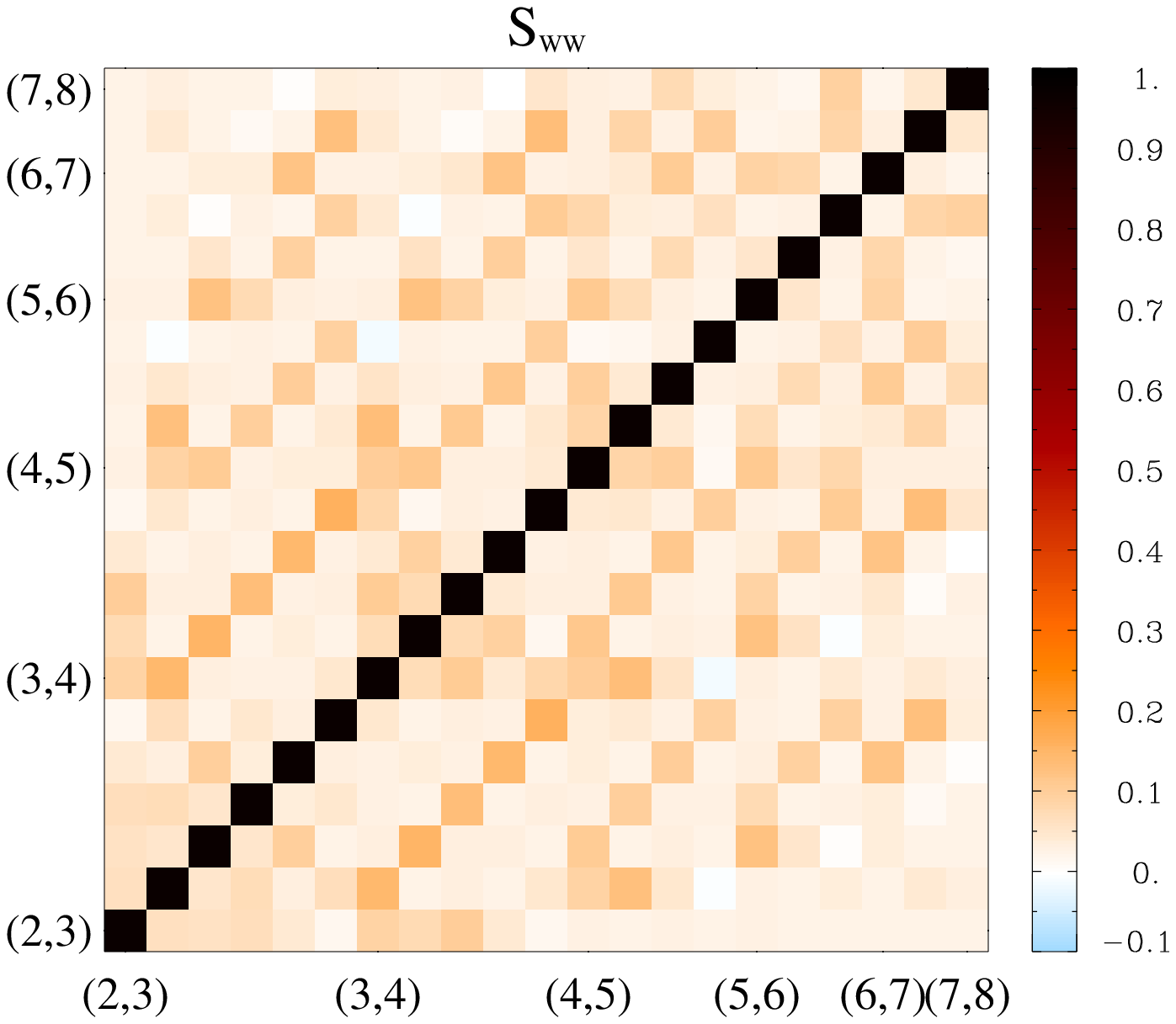,width=0.4\linewidth} 
\end{tabular}

\caption{ The correlation matrices of the multipole vectors statistics
($S_{\rm vv}$, $S_{\rm vw}$, $S_{\rm wv}$ and $S_{\rm ww}$ from the
upper to the lower row, respectively) for the universe with the T228
topology without the ISW (left column) and with ISW effect (right
column).  Pairs of multipoles are ordered in ascending order of the
multipoles i.e., $(2,3),\, (2,4), \ldots,\, (2,8),\, (3,4),\,
(3,5),\ldots, (7,8)$.  } \label{fig:corr_t228}
\end{figure*}

\begin{figure*}
\begin{tabular}{cc}
 \epsfig{file=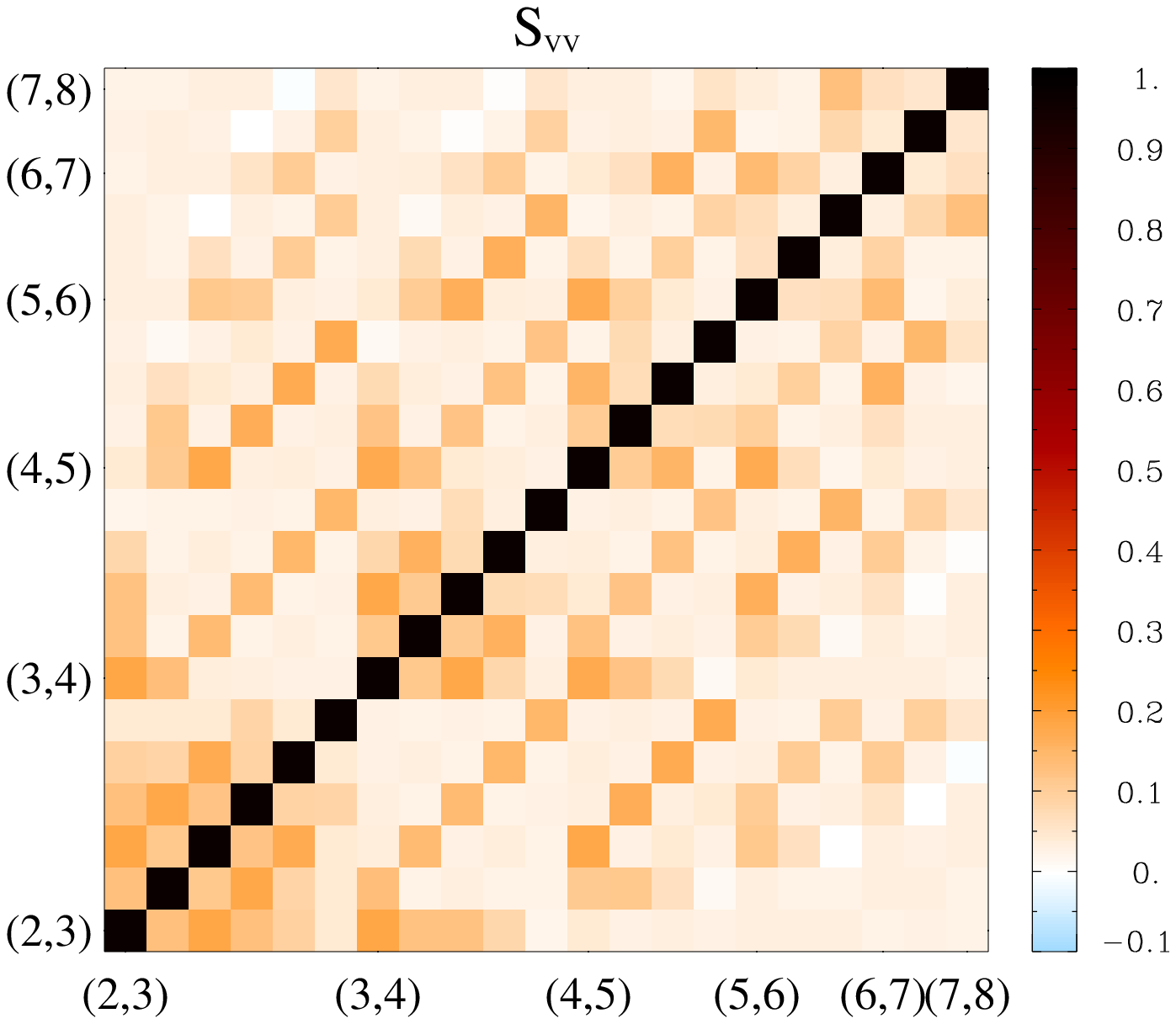,width=0.4\linewidth} &
 \epsfig{file=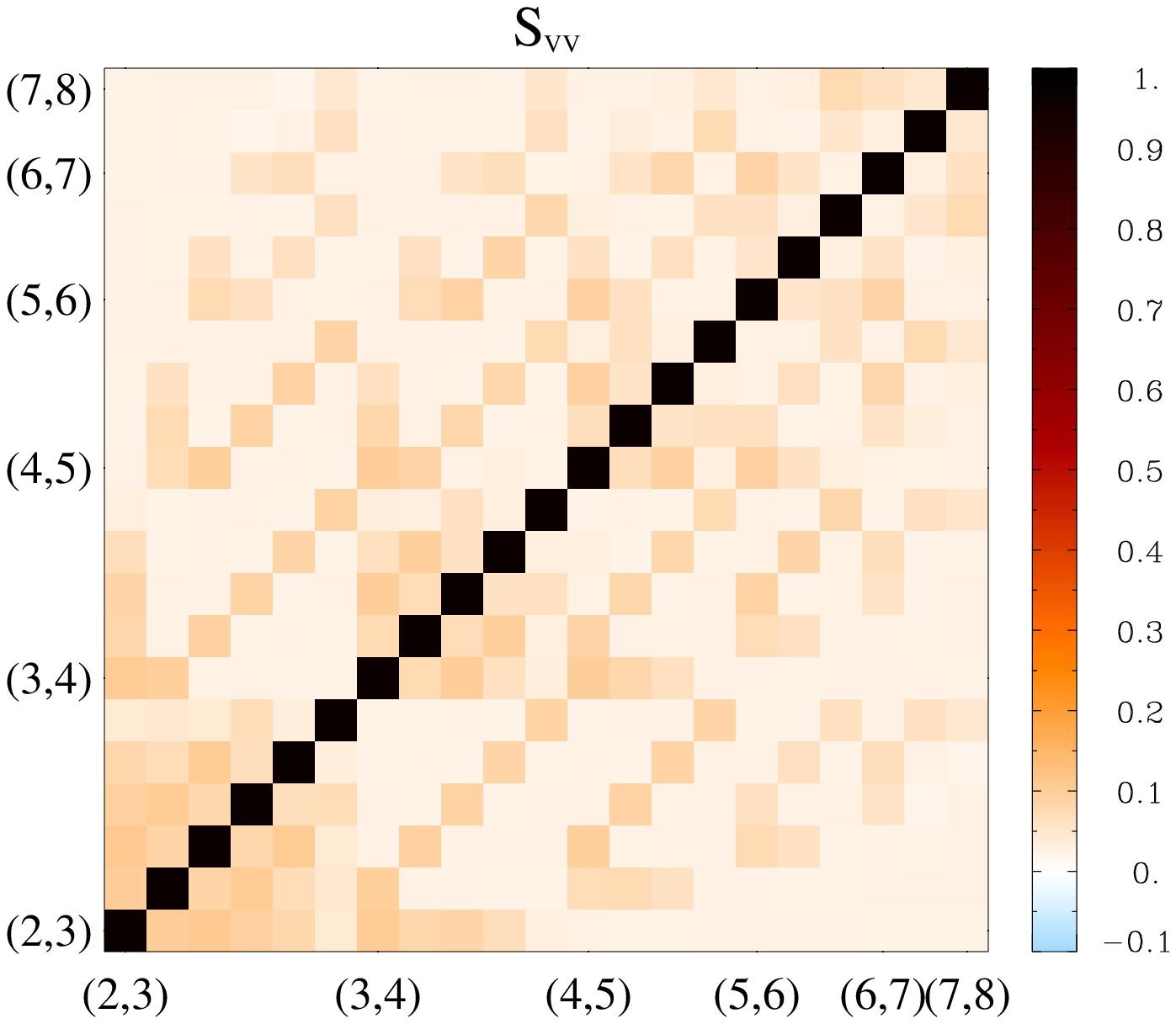,width=0.4\linewidth} \\
 \epsfig{file=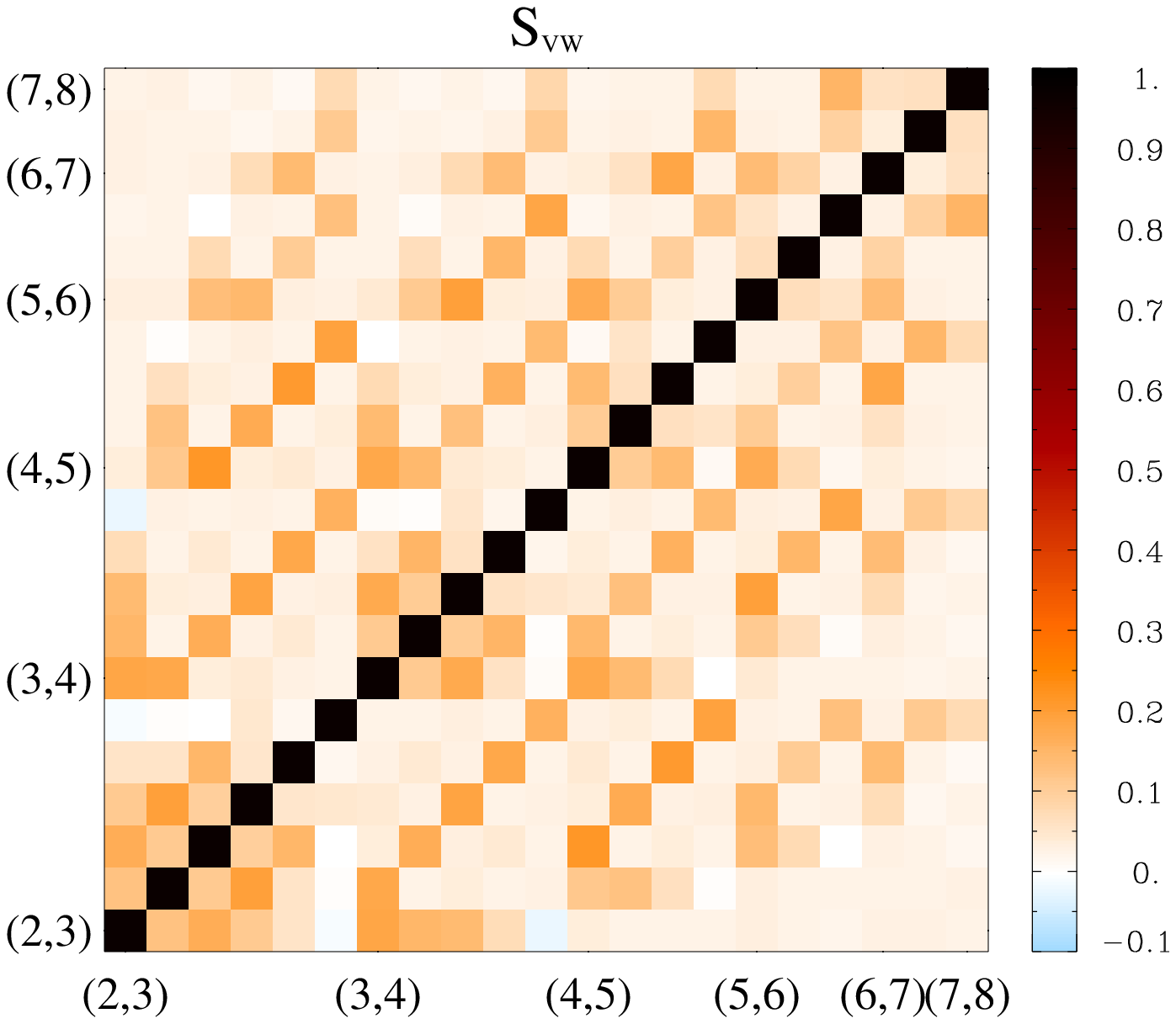,width=0.4\linewidth} &
 \epsfig{file=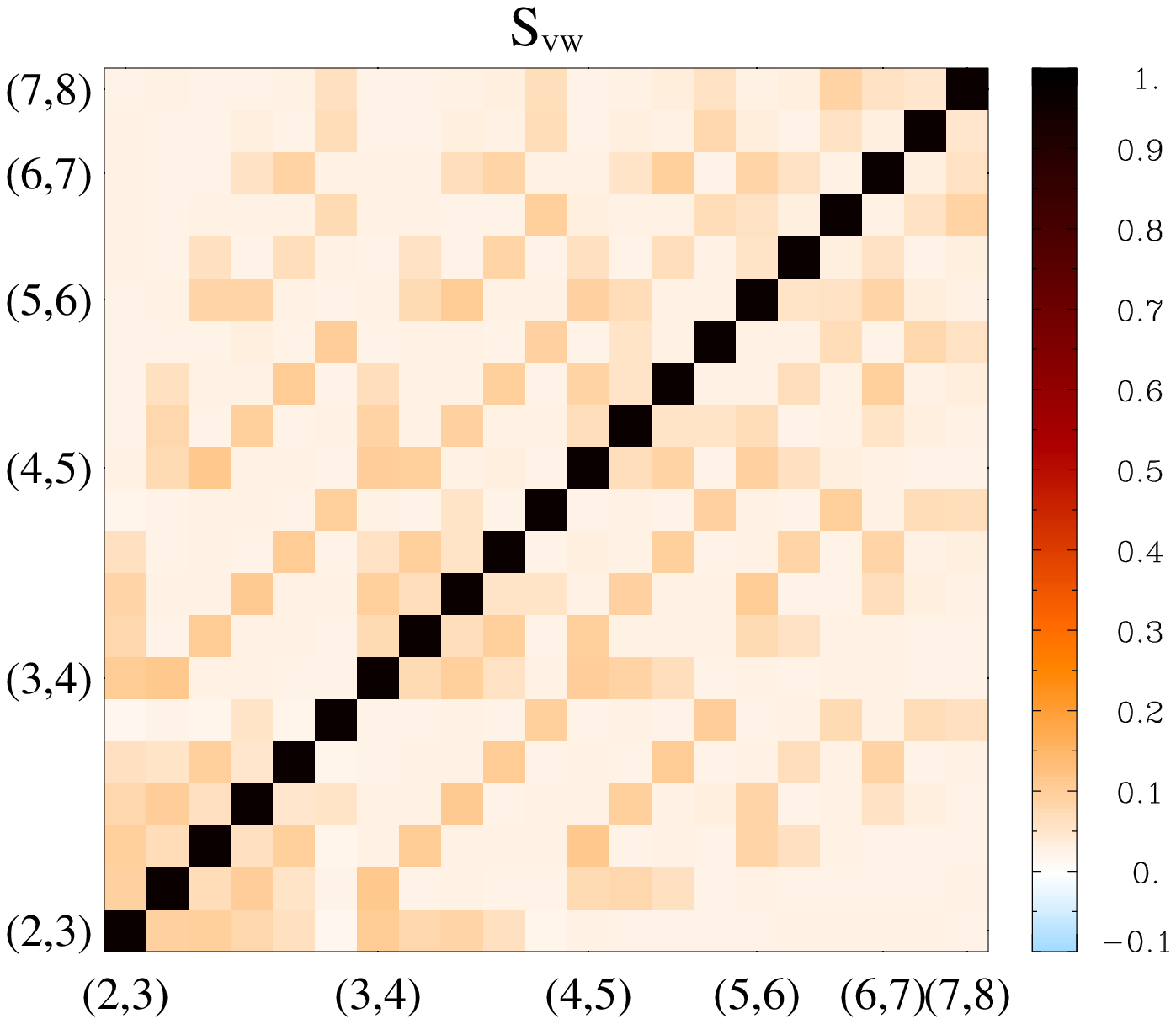,width=0.4\linewidth} \\
 \epsfig{file=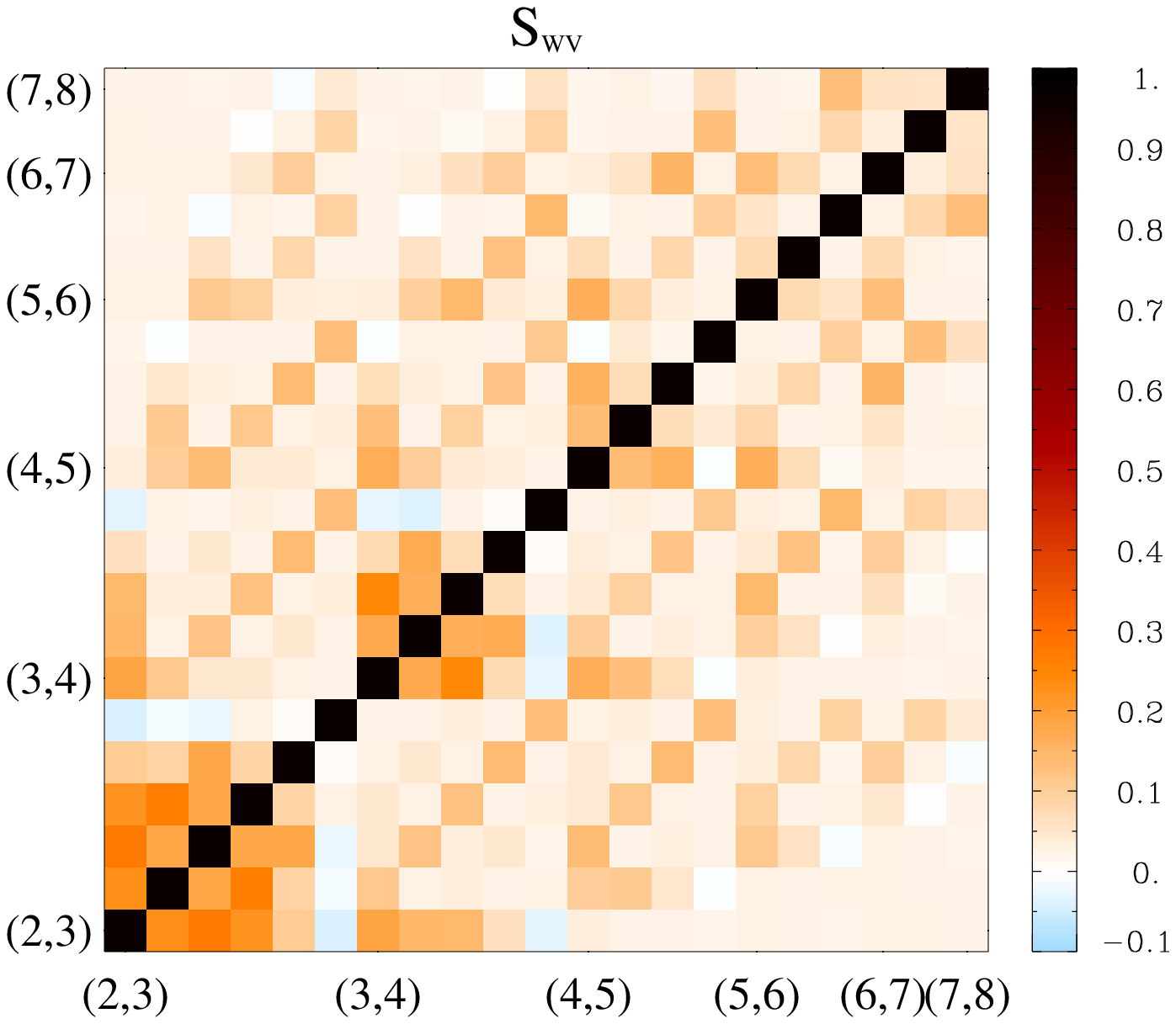,width=0.4\linewidth} &
 \epsfig{file=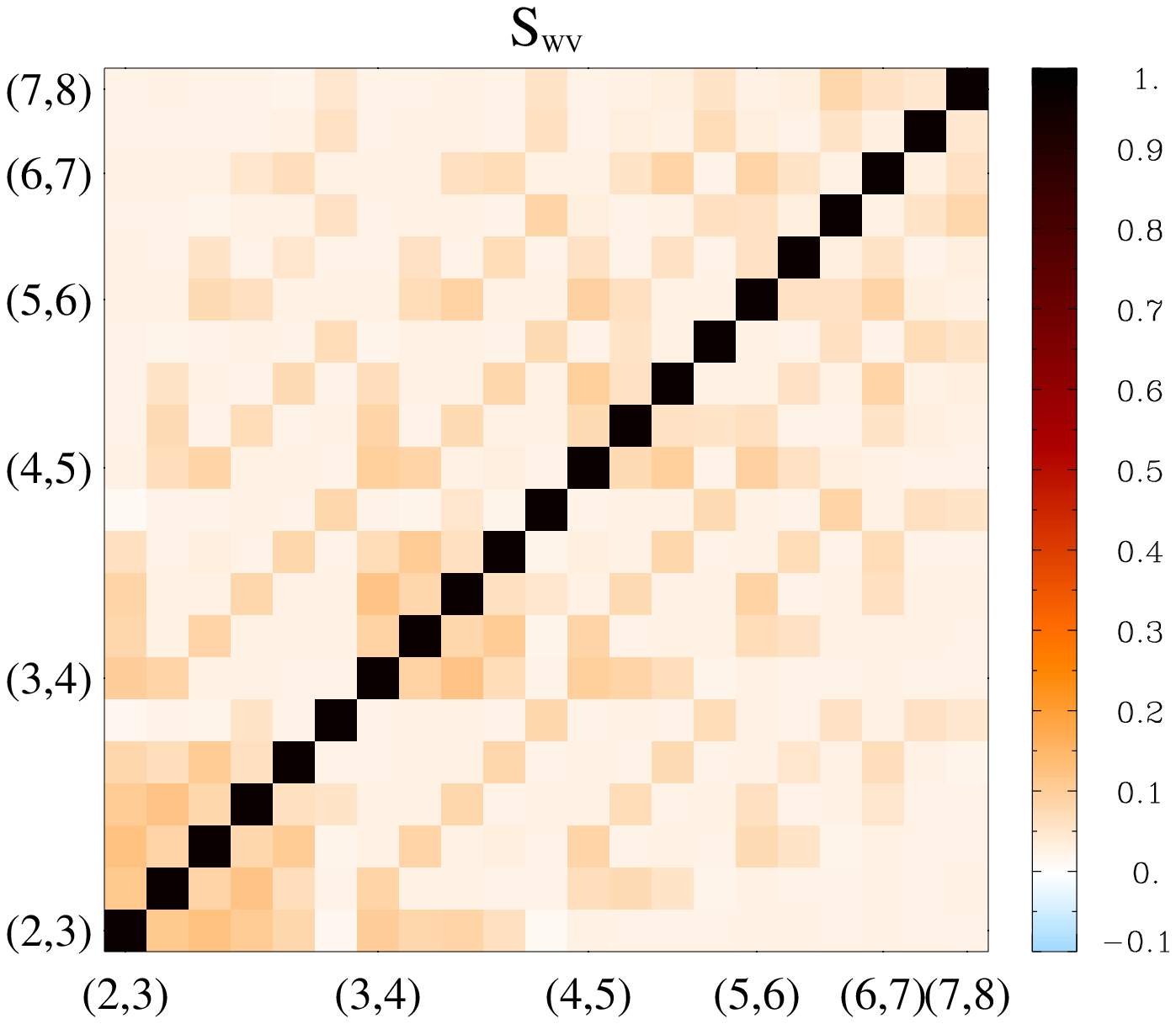,width=0.4\linewidth} \\
 \epsfig{file=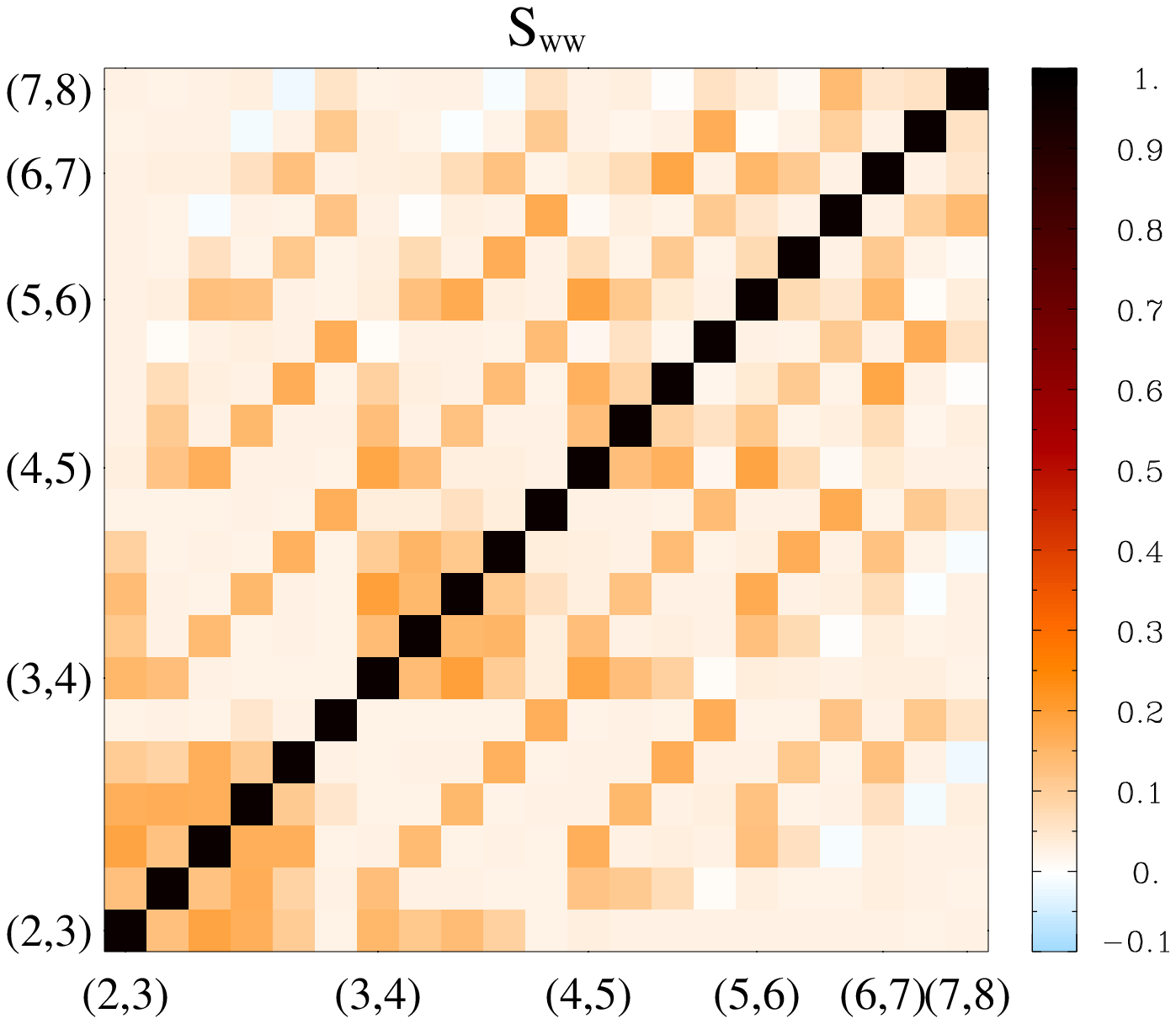,width=0.4\linewidth} &
 \epsfig{file=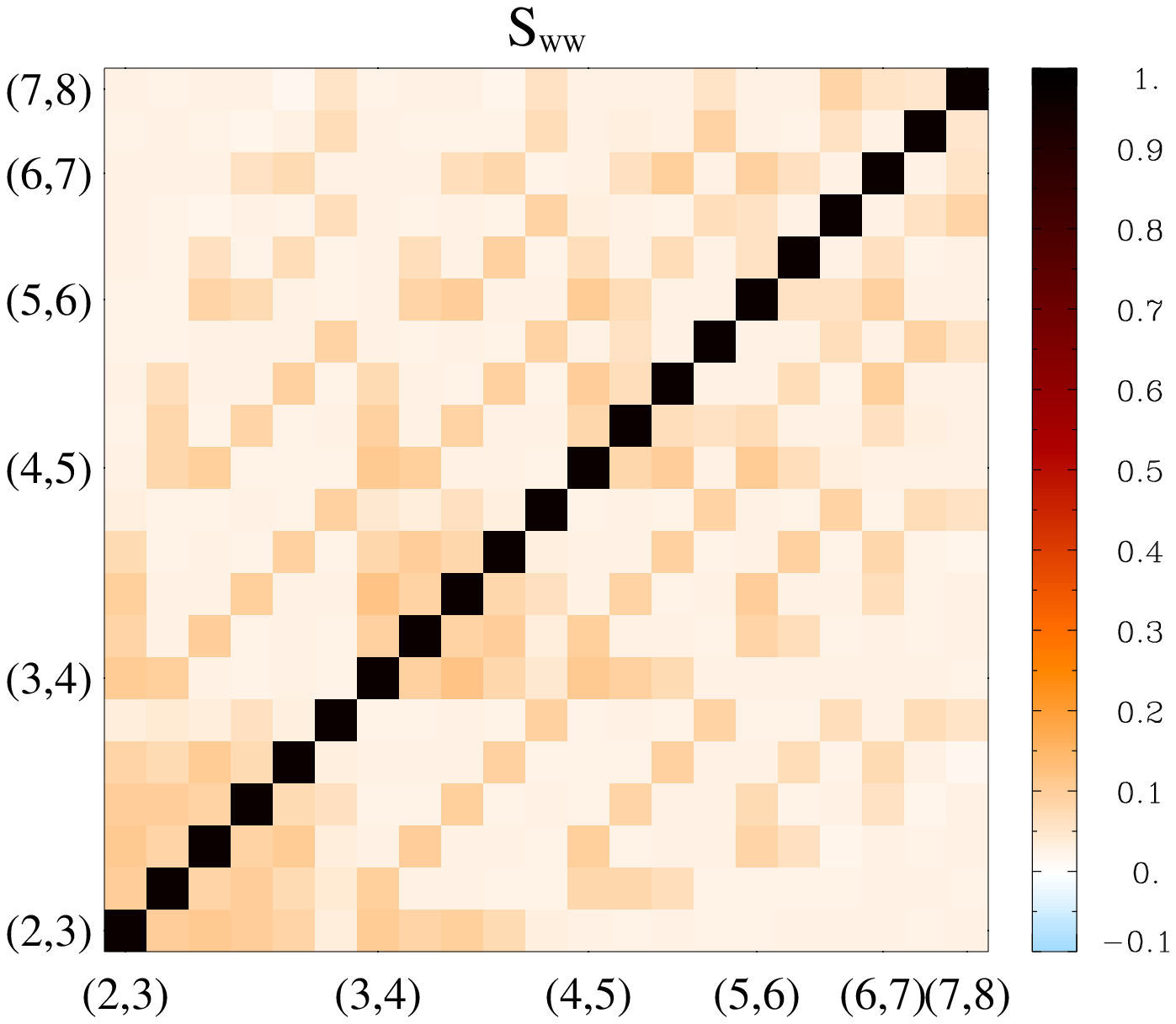,width=0.4\linewidth} 
\end{tabular}

\caption{ Correlation matrices of the multipole vectors statistics
($S_{\rm vv}$, $S_{\rm vw}$, $S_{\rm wv}$ and $S_{\rm ww}$ from the
upper to the lower row, respectively) for the universe with the T882
topology without the ISW (left column) and with ISW effect (right
column).  Pairs of multipoles are ordered in ascending order of the
multipoles i.e., $(2,3),\, (2,4), \ldots,\, (2,8),\, (3,4),\,
(3,5),\ldots, (7,8)$.  } \label{fig:corr_t882}
\end{figure*}

To quantify how well a universe with different topologies fits the
data, we estimated the probabilities of obtaining values for the statistics
smaller than the data.  
Since in a multi-connected topology the statistics for different pairs of the multipoles are
correlated with each other (see correlation matrices on
\fig\ref{fig:corr_t228} and \fig\ref{fig:corr_t882}), to correctly
estimate how well the theoretical models fit we need to consider a joint
PDF of the statistics for a few pairs of the multipoles.

From a set of $10^6$ simulations we were able to estimate the joint
distribution for up to 4 pairs of the statistics. We chose pairs which
are the most strongly correlated with each other. The correlation
matrices, shown in \fig\ref{fig:corr_t228} and \fig\ref{fig:corr_t882},
reveal significant correlations between the statistics where one of the
multipole in the pair is quadrupole or octopole. The biggest
correlations, up to 0.4 for the maps without the ISW effect and the T228
topology, are between the statistics for the pairs $(2,\ell)$ and
$(3,\ell)$, where $\ell$ takes value in the range $\ell \in [4,8]$. They
are consequence of the alignment of the multipole vectors in the direction of
the longer sides of the fundamental domains. As was shown in
\sect\ref{sec:distribution}, in the case of the quadrupole and octopole the
alignment is stronger than for higher order multipoles. Let us emphasize
that these alignments are somehow unrelated to the multipole coefficient
correlation matrix $\left<a_{\ell m} a^*_{\ell'm'} \right>$. Actually,
for parity reasons, coefficients of the second and third multipoles are
statistically uncorrelated for any torus (i.e.,
$\left<a_{2m}a^*_{3m'}\right> = 0$ for all $m$, $m'$). However, because
of the fundamental domain shape, both the quadrupole and octopole are led to
show up configurations which feel the elongated or flattened direction
of our torus, and as a consequence, their corresponding multipole vector
do show some alignment, even though the multipole coefficients
themselves are uncorrelated. It happens that here, the correlations
caused by the alignment dominate over correlations between multipoles
sharing the same parity, like for instance $\ell=2$ and $\ell=4$, which
could be expected to be quite strong.

We decided to analyse triplets of the statistics for the pairs $(2,3)$,
$(2,\ell)$ and $(3,\ell)$ (hereafter referred as the $\ell$-th triplet)
as well quartets -- for the pairs $(2,3)$, $(2,4)$, $(2,5)$, $(2,6)$
(hereafter referred as the quadrupole quartet) and $(3,4)$, $(3,5)$,
$(3,6)$, $(3,7)$ (hereafter referred as the octopole quartet). We chose
also a quartet consisting of the pairs $(2,3)$, $(2,4)$, $(3,4)$, $(3,7)$
for which the observed values of the statistics deviate substantially
from the predictions of the simply-connected universe. The probability of
obtaining values at most as extreme as for the data under the null
hypothesis (i.e., simply-connected universe) should be unusually big for
this particular quartet.

The probabilities for the triplets and quartets of the statistics under
different hypotheses about the topology are presented in \tab\ref{tab:prob1}
and \tab\ref{tab:prob2}, respectively. In the case of the statistics with
cross products of the multipole vectors, we present probabilities
corresponding to both normalised and unnormalised cross products.

With few exceptions, the estimated probabilities indicate
that universes with a multi-connected topology are a better fit to the data, 
irrespective of whether the ISW effect is taken into account.
The T882 topology is preferred by the triplets $\{(2,3),\ (2,4),\
(3,4)\}$ and $\{(2,3),\ (2,7),\ (3,7)\}$ as well the octopole and
$\{(2,3),\ (2,4),\ (3,4),\ (3,7)\}$ quartets for all types of
statistics. The preference for the T228 topology by the data depends
more strongly on the type of the statistic and presence of the ISW effect.

It is interesting that a few of the triplets of the $S_{\rm vv}$ and $S_{\rm ww}$
statistics, for example $\{(2,3),\ (2,4),\ (3,4)\}$ or $\{(2,3),\
(2,5),\ (3,5)\}$, more strongly favour the multi-connected topology
for maps with the ISW effect included than excluded.
This is the case for both
of the considered topologies. This rather contradicts expectations  -- 
the ISW effect tends to dilute the signatures of multiconnected topology,
therefore the probabilities should be higher for the maps without
the ISW effect, like for the simply-connected universe.  It may indicate
that the observed relative relations between the vectors for these
multipoles do not fit well enough to the characteristic distribution of
the vectors for those topologies shown in
\sect\ref{sec:distribution}. They only follow the directions preferred
by the topologies.

The probabilities for the quadrupole quartet and the statistics with the
ISW effect and normalised cross products, except for the $S_{\rm ww}$
statistic, show a stronger preference for the T228 topology
as opposed to the T882 one. However, for the statistics with unnormalised
cross products the situation is reversed -- the data prefer rather the
T882 topology.  On the other hand, for the octopole quartet the
statistics with normalised cross product prefers the T882 topology
while with unnormalised cross products, expect for the $S^u_{\rm wv}$
statistic, -- the T228 one is supported.

In general the probabilities for the $S_{\rm ww}^u$ statistics are
higher than for the statistics with the normalised cross products
$S_{\rm ww}$. For the multi-connected universe they include a few cases
with values as extreme as $99\%$ or even $100\%$ for the $\{(2,3),\ (2,6),\
(3,6)\}$ triplet. In contrast to the $S_{\rm ww}$ statistics, they
favour a simply-connected universe compared to a
multi-connected one.

It is worth emphasizing that, for any of the analysed triplets and
quartets of the statistics, the computed probabilities do not clearly
rule out the simply-connected universe. For the statistics with the
normalised cross products the $\{(2,3),\ (2,4),\ (3,4),\ (3,7)\}$
quartet have the biggest probability, up to 94\%. To a large extent,
the correlations of the $\{(2,3),\ (2,4),\ (3,4)\}$
triplet are responsible for this behaviour. In this case, similarly large 
values of the probability are seen. For the rest of the triplets and quartets of
this type, we do not observe any unusually extreme statistical values.

For the statistics with the unnormalised cross products the most extreme
probabilities are higher, up to 97\% for the $\{(2,3),\ (2,6),\ (3,6)\}$
triplet and 96\% for the quadrupole quartet.

\begin{table}
\begin{center}
\caption{ The probabilities, $P_S$, of obtaining values of the
statistics, $S_{\rm vv}$, $S_{\rm vw}$, $S_{\rm wv}$, $S_{\rm ww}$,
$S_{\rm vw}^u$, $S_{\rm wv}^u$, $S_{\rm ww}^u$, respectively, smaller
than the observed values of the statistics for the ILC \emph{WMAP}
5 years map. SC denotes the probabilities for the simply-connected
topology, T228 and T228$^\ast$ -- for the CMB maps of the universe with
the T228 topology without and with the ISW effect,
respectively. Similarly, T882 and T882$^\ast$ denote the probabilities
for the universe with the T882 topology.  } \label{tab:prob1} For pairs
of multipoles $(\ell_1, \ell_2) = \{(2,3),\ (2,4),\ (3,4)\}$\\
\begin{tabular}{|c|ccccc|}\hline
 & SC & T228 & T228$^\ast$ & T882 & T882$^\ast$ \\ \hline
 $P_{S_{\rm vv}}$ & 90\% & 92\% & 89\% & 87\% & 88\%  \\
 $P_{S_{\rm vw}}$ & 92\% & 69\% & 82\% & 74\% & 88\%  \\
 $P_{S_{\rm wv}}$ & 94\% & 83\% & 90\% & 87\% & 93\%  \\
 $P_{S_{\rm ww}}$ & 94\% & 96\% & 97\% & 84\% & 88\%  \\
 $P_{S_{\rm vw}^u}$ & 73\% & 27\% & 47\% & 65\% & 63\%  \\
 $P_{S_{\rm wv}^u}$ & 95\% & 99\% & 97\% & 93\% & 94\%  \\
 $P_{S_{\rm ww}^u}$ & 95\% & 99\% & 98\% & 94\% & 93\%  \\
\hline				  		      	    		  
\end{tabular}			  		      	    	  
\\		
For pairs of multipoles $(\ell_1, \ell_2) = \{(2,3),\ (2,5),\ (3,5)\}$\\
\begin{tabular}{|c|ccccc|}\hline
 & SC & T228 & T228$^\ast$ & T882 & T882$^\ast$ \\ \hline
 $P_{S_{\rm vv}}$ & 51\% & 44\% & 17\% & 44\% & 33\%  \\
 $P_{S_{\rm vw}}$ & 69\% & 8\%  & 29\% & 36\% & 54\%  \\
 $P_{S_{\rm wv}}$ & 71\% & 12\% & 52\% & 26\% & 63\%  \\
 $P_{S_{\rm ww}}$ & 80\% & 76\% & 79\% & 79\% & 75\%  \\
 $P_{S_{\rm vw}^u}$ & 74\% & 18\% & 48\% & 49\% & 52\%  \\
 $P_{S_{\rm wv}^u}$ & 67\% & 68\% & 41\% & 42\% & 52\%  \\
 $P_{S_{\rm ww}^u}$ & 87\% & 96\% & 98\% & 86\% & 94\%  \\
\hline				  		      	    		  
\end{tabular}			  		      	    	  
\\
For pairs of multipoles $(\ell_1, \ell_2) = \{(2,3),\ (2,6),\ (3,6)\}$\\	
\begin{tabular}{|c|ccccc|}\hline
 & SC & T228 & T228$^\ast$ & T882 & T882$^\ast$ \\ \hline
 $P_{S_{\rm vv}}$ & 49\% & 55\% & 26\% & 29\% & 55\%  \\
 $P_{S_{\rm vw}}$ & 86\% & 67\% & 73\% & 63\% & 75\%  \\
 $P_{S_{\rm wv}}$ & 88\% & 20\% & 48\% & 48\% & 69\%  \\
 $P_{S_{\rm ww}}$ & 91\% & 88\% & 88\% & 96\% & 89\%  \\
 $P_{S_{\rm vw}^u}$ & 85\% & 25\% & 69\% & 64\% & 67\%  \\
 $P_{S_{\rm wv}^u}$ & 65\% & 56\% & 43\% & 27\% & 47\%  \\
 $P_{S_{\rm ww}^u}$ & 97\% & 100\% & 96\% & 99\% & 98\%   \\
\hline				  		      	    		  
\end{tabular}			  		      	    	  
\\
For pairs of multipoles $(\ell_1, \ell_2) = \{(2,3),\ (2,7),\ (3,7)\}$\\		
\begin{tabular}{|c|ccccc|}\hline
 & SC & T228 & T228$^\ast$ & T882 & T882$^\ast$  \\ \hline
 $P_{S_{\rm vv}}$ & 68\% & 40\% & 52\% & 18\% & 40\%  \\
 $P_{S_{\rm vw}}$ & 80\% & 28\% & 39\% & 52\% & 66\%  \\
 $P_{S_{\rm wv}}$ & 82\% & 30\% & 62\% & 62\% & 60\%  \\
 $P_{S_{\rm ww}}$ & 92\% & 92\% & 88\% & 76\% & 78\%  \\
 $P_{S_{\rm vw}^u}$ & 82\% & 28\% & 44\% & 45\% & 70\%  \\
 $P_{S_{\rm wv}^u}$ & 71\% & 30\% & 37\% & 24\% & 52\%  \\
 $P_{S_{\rm ww}^u}$ & 92\% & 99\% & 94\% & 97\% & 75\%  \\
\hline				  		      	    		  
\end{tabular}			  		      	    	  	
\\
For pairs of multipoles $(\ell_1, \ell_2) = \{(2,3),\ (2,8),\ (3,8)\}$\\
\begin{tabular}{|c|ccccc|}\hline
 & SC & T228 & T228$^\ast$ & T882 & T882$^\ast$ \\ \hline
 $P_{S_{\rm vv}}$ & 51\% & 33\% & 34\% & 25\% & 55\%  \\
 $P_{S_{\rm vw}}$ & 77\% & 8\%  & 50\% & 51\% & 61\% \\
 $P_{S_{\rm wv}}$ & 80\% & 27\% & 49\% & 41\% & 64\%  \\
 $P_{S_{\rm ww}}$ & 80\% & 88\% & 76\% & 74\% & 69\%  \\
 $P_{S_{\rm vw}^u}$ & 81\% & 12\% & 41\% & 37\% & 62\%  \\
 $P_{S_{\rm wv}^u}$ & 70\% & 66\% & 49\% & 26\% & 49\%  \\
 $P_{S_{\rm ww}^u}$ & 80\% & 93\% & 93\% & 80\% & 87\%  \\
\hline				  		      	    		  
\end{tabular}			  		      	    	  
\end{center}
\end{table}

\begin{table}
\begin{center}
\caption{
The same as in Table \ref{tab:prob1} but for the quartets of the statistics.
}
\label{tab:prob2}
For pairs of multipoles $(\ell_1, \ell_2) = \{(2,3),\ (2,4),\ (2,5),\ (2,6)\}$\\	
\begin{tabular}{|c|ccccc|}\hline
 & SC & T228 & T228$^\ast$ & T882 & T882$^\ast$ \\ \hline
 $P_{S_{\rm vv}}$ & 68\% & 53\% & 43\% & 66\% & 74\%  \\
 $P_{S_{\rm vw}}$ & 88\% & 74\% & 83\% & 66\% & 86\%  \\
 $P_{S_{\rm wv}}$ & 88\% & 66\% & 81\% & 78\% & 91\%  \\
 $P_{S_{\rm ww}}$ & 89\% & 93\% & 93\% & 80\% & 85\%  \\
 $P_{S_{\rm vw}^u}$ & 65\% & 29\% & 58\% & 66\% & 56\%  \\
 $P_{S_{\rm wv}^u}$ & 86\% & 83\% & 81\% & 76\% & 71\%  \\
 $P_{S_{\rm ww}^u}$ & 96\% & 99\% & 99\% & 85\% & 93\%  \\
\hline 
\end{tabular}					      				    
\\
For pairs of multipoles $(\ell_1, \ell_2) = \{(3,4),\ (3,5),\ (3,6),\ (3,7)\}$\\
\begin{tabular}{|c|ccccc|}\hline
 & SC & T228 &  T228$^\ast$ & T882 & T882$^\ast$ \\ \hline
 $P_{S_{\rm vv}}$ & 88\% & 59\% & 88\% & 79\% & 79\%  \\
 $P_{S_{\rm vw}}$ & 52\% & 48\% & 55\% & 39\% & 51\%  \\
 $P_{S_{\rm wv}}$ & 88\% & 83\% & 78\% & 77\% & 78\%  \\
 $P_{S_{\rm ww}}$ & 88\% & 92\% & 89\% & 67\% & 76\%  \\
 $P_{S_{\rm vw}^u}$ & 34\% & 15\% & 18\% & 17\% & 29\%  \\
 $P_{S_{\rm wv}^u}$ & 82\% & 94\% & 87\% & 68\% & 77\%  \\
 $P_{S_{\rm ww}^u}$ & 80\% & 85\% & 73\% & 70\% & 80\%  \\
\hline
\end{tabular}			  		      	    	  
\\
For pairs of multipoles $(\ell_1, \ell_2) = \{(2,3),\ (2,4),\ (3,4),\ (3,7)\}$\\		
\begin{tabular}{|c|ccccc|}\hline
 & SC & T228 & T228$^\ast$ & T882 & T882$^\ast$ \\ \hline
 $P_{S_{\rm vv}}$ & 92\% & 87\% & 90\% & 83\% & 84\%  \\
 $P_{S_{\rm vw}}$ & 87\% & 62\% & 76\% & 70\% & 81\%  \\
 $P_{S_{\rm wv}}$ & 92\% & 83\% & 86\% & 86\% & 89\%  \\
 $P_{S_{\rm ww}}$ & 94\% & 95\% & 95\% & 78\% & 83\%  \\
 $P_{S_{\rm vw}^u}$ & 69\% & 22\% & 42\% & 57\% & 62\%  \\
 $P_{S_{\rm wv}^u}$ & 92\% & 97\% & 95\% & 84\% & 86\%  \\
 $P_{S_{\rm ww}^u}$ & 91\% & 99\% & 94\% & 89\% & 91\%  \\
\hline
\end{tabular}
\end{center}
\end{table}

Before we draw any conclusions from these results, one needs to point
out that the probabilities shown in \tab\ref{tab:prob1} and
\tab\ref{tab:prob2} are merely an approximation to the true values.
Because of the limitation imposed on the number of dimensions that we
could consider, we did not study the joint probability distribution
function defined for more than 4-dimensional space and did not take into
account other correlations, clearly visible in \fig\ref{fig:corr_t228}
and \fig\ref{fig:corr_t882}, between the statistics.  One needs also to
remember that the sets of multipoles pairs considered were selected on 
the basis of 
a posteriori knowledge concerning the values of the statistics for the
data. This could also influence our inferences regarding the CMB isotropy.

\section{Summary} \label{sec:conclusions}

We have studied signatures of a multiconnected topology of the universe via the
distribution of multipole vectors on the sky and the PDF of their
statistics. We considered topologies for a flat universe, where a
fundamental domain was a rectangular prism with dimensions $L_x = L_y = 2
R_H$, $L_z= 8 R_H$ (elongated case) and $L_x = L_y = 8 R_H$, $L_z= 2
R_H$ (flattened case). In both cases, only the shortest dimension of the
prism is smaller than diameter of the last scattering
surface. Unsurprisingly, the distribution of the multipole vectors on
the sky, especially for the quadrupole, reflects symmetries of the 
fundamental cell specific to the topology under consideration . The multipole
vectors are aligned along the longer sides of the fundamental
cell, but the alignment is more pronounced for the quadrupole and
octopole. As the ISW effect is caused by the evolution of structures
close to the observer, it significantly diminishes the evidence of a
multiconnected topology, so that the vectors follow instead
the distribution of the power on the CMB
maps. For some of the statistics, the joint PDF for a few pairs of the multipoles indicates
that the data is better fitted by a model of the universe with
a multiconnected topology. However, one can also find statistics
for which certain pairs support the alternative conclusions. 
A more quantitative assessment indicates that the data do indeed
slightly prefer the multiconnected topology. 
Nevertheless, the most significant values of the probabilities
for the simply-connected universe, i.e.\ 94\% for the statistics with
normalised cross products and 97\% for the statistics with unnormalised
products, show that this hypothesis is not clearly ruled out. We also found that
the multipole vectors statistics are not very sensitive to the
signatures of the 3-torus topology if the shorter dimension of the
fundamental domain is comparable to the observable diameter of the
universe.

These results have a rather preliminary character.  There is a need to
carefully evaluate the influence of the contamination from the Galactic
foreground on detection of the topological signatures. It is known that
the \emph{WMAP} ILC map used in our studies to some extent is
contaminated by foreground residuals. The maps for the Q, V and
W-bands with the masked Galactic plane and corrected foreground are more
credible for the cosmological analysis. However, application of the
mask breaks the statistical isotropy of the maps. Then, we have
to consider all possible orientations of the fundamental domain with
respect to the mask, which makes the analysis computationally much
tougher.

Extension of the studies is also seriously limited by the fact that we
do not have any analytical expression for the PDF of the multipole
vectors statistics even for the simply-connected universe. So far only
the PDF of the multipole vectors for the simply-connected universe was
given \citep{dennis:2008}. Thus, we had to base our analysis on
simulations. The correlations between different pairs of the statistics
present for multi-connected universes introduces the necessity of a
multivariate analysis of the joint PDF to allow a correct estimation of the
probabilities. However, using $10^6$ simulations this could only be
achieved up to a four dimensional subspace.

\section*{Acknowledgements}
We thank the anonymous referee for useful suggestions that substantially
improved the article and Tony Banday for a careful reading of
the manuscript. We acknowledge the use 
of CMBFAST \citep{cmbfast}. Some of the results 
in this paper have been derived using the
HEALPix\footnote{http://healpix.jpl.nasa.gov} \citep{healpix} software
and analysis package.  We acknowledge the use of the Legacy Archive for
Microwave Background Data Analysis\footnote{http://lambda.gsfc.nasa.gov}
(LAMBDA).  Support for LAMBDA is provided by the NASA Office of Space
Science.  PB thanks the Agence Nationale de la Recherche grant
ANR-05-BLAN-0289-01 for support. 



\begin{appendix} 
\section{The multipole vectors for quadrupole} \label{sec:appendix}

In case of quadrupole it is possible to show explicitly relations 
between the spherical harmonics coefficients and the multipole vectors.
Quadrupole may be regarded as 3$\times$3 symmetric traceless matrix 
$Q_{ij}$  such that
\begin{eqnarray} \label{} 
T_2(\hat{\mathbf{e}}) &=& \sum_{m=-2}^2 a_{2m} Y_{2m}(\hat{\mathbf{e}}) =
 A^{(2)}\, \left[(\hat{\mathbf{v}}^{(2,1)}
 \cdot \hat{\mathbf{e}}) (\hat{\mathbf{v}}^{(2,2)}
 \cdot \hat{\mathbf{e}}) + \right. \nonumber \\
 & & \left. - {1 \over 3} \hat{\mathbf{v}}^{(2,1)} \cdot
 \hat{\mathbf{v}}^{(2,2)}\right] = Q_{ij} x^i x^j \ .
\end{eqnarray}
Considering Cartesian coordinates $x^i = (z,x,y)$ with constraint $x^2+y^2+z^2=1$
the $Q_{ij}$ matrix takes the form
\begin{equation}
Q_{ij} = \sqrt{{5 \over 16 \pi}} \left(
\begin{array}{ccc}
{2\, a_{20} \over \sqrt{3}} & -\sqrt{2}\, a_{21}^{\rm Re} & \sqrt{2}\,
 a_{21}^{\rm Im} \\
-\sqrt{2}\, a_{21}^{\rm Re} & -{a_{20} \over \sqrt{3}} + \sqrt{2}\,
 a_{22}^{\rm Re} & -\sqrt{2}\, a_{22}^{\rm Im} \\
\sqrt{2}\, a_{21}^{\rm Im} & -\sqrt{2}\, a_{22}^{\rm Im} & -{a_{20} \over \sqrt{3}} - \sqrt{2}\,
 a_{22}^{\rm Re} 
\end{array}
\right) \ .
\end{equation}
As pointed out by \cite{land:2005} and \citet{copi:2006} eigenvectors of
$Q$ are related to the multipole vectors by
\begin{eqnarray} 
\mathbf{V}^1 &\propto& \hat{\mathbf{v}}^{(2,1)} +
 \hat{\mathbf{v}}^{(2,2)} \label{eqn:eigenvec_multipole1} \\
\mathbf{V}^2 &\propto& \hat{\mathbf{v}}^{(2,1)} -
 \hat{\mathbf{v}}^{(2,2)} \label{eqn:eigenvec_multipole2} \\
\mathbf{V}^3 &\propto& \hat{\mathbf{v}}^{(2,1)} \times \hat{\mathbf{v}}^{(2,2)}
\end{eqnarray}
with corresponding eigenvalues $\lambda_1 = A^{(2)}\, (X + 3)/6$,
\mbox{$\lambda_2 = A^{(2)}\, (X - 3)/6$}, $\lambda_3 = -A^{(2)}\, X/3$,
where
\begin{equation} \label{eqn:dot_prod}
X \equiv \hat{\mathbf{v}}^{(2,1)} \cdot \hat{\mathbf{v}}^{(2,2)} = 3 (\lambda_1 +
\lambda_2)/|\lambda_1 - \lambda_2| \ .
\end{equation} 

On the other hand, eigenvector for a given eigenvalue $\lambda$ can be
expressed in terms of the elements of the $Q$ matrix as 
\begin{eqnarray} 
\hat{V}_x &=& {1 \over N} \left[ -Q_{zx}\, (Q_{xx}+Q_{zz}+\lambda) - Q_{xy}\, Q_{zy}\right] \ , \label{eqn:eigenvec_matrix1} \\
\hat{V}_y &=& {1 \over N} \left[ Q_{zy}\, (Q_{xx}-\lambda) - Q_{xy}\, Q_{zx}\right] \ , \label{eqn:eigenvec_matrix2}\\
\hat{V}_z &=& {1 \over N} \left[ (Q_{xx} - \lambda) (Q_{xx} + Q_{zz} + \lambda) + Q_{xy}^2
 \right]\ ,  \label{eqn:eigenvec_matrix3}
\end{eqnarray}
where $N$ is normalisation factor chosen such that
\mbox{$|\hat{\mathbf{V}}|=1$}. The eigenvalues are roots of the
characteristic polynomial $\lambda^3 - p\, \lambda - \det Q = 0$, where
$p = Q_{xx}^2 + Q_{zz}^2 + Q_{zy}^2 + Q_{xy}^2+ Q_{zx}^2 + Q_{xx}
Q_{zz}$.

Thus, using (\ref{eqn:eigenvec_multipole1}),
(\ref{eqn:eigenvec_multipole2}) and (\ref{eqn:eigenvec_matrix1}),
(\ref{eqn:eigenvec_matrix2}), (\ref{eqn:eigenvec_matrix3}) one obtains
direct relation between the spherical harmonics coefficients and the
multipole vectors
\begin{eqnarray}
\hat{\mathbf{v}}^{(2,1)} &=& W_1\, \hat{\mathbf{V}}^1 + W_2\, \hat{\mathbf{V}}^2 \ , \\
\hat{\mathbf{v}}^{(2,2)} &=& W_1\, \hat{\mathbf{V}}^1 - W_2\, \hat{\mathbf{V}}^2 \ ,
\end{eqnarray}
where weights $W_1=\sqrt{(1+X)/2}$ and $W_2=\sqrt{(1-X)/2}$
ensure correct normalisation of the multipole vectors
$|\hat{\mathbf{v}}^{(2,1)}| = |\hat{\mathbf{v}}^{(2,2)}| = 1$ and
relation (\ref{eqn:dot_prod}).

\end{appendix}



\begin{thebibliography}{}

\bibitem[Adams \& Shapiro(2001)]{adams:2001}
Adams C., Shapiro J., 2001, Am.~Sci., 89, 443

\bibitem[Aurich, Lustig \& Steiner(2005a)]{aurich:2005a}
Aurich R., Lustig S., Steiner F., 2005a, CQGra, 22, 2061

\bibitem[Aurich, Lustig \& Steiner(2005b)]{aurich:2005b} 
Aurich R., Lustig S., Steiner F., 2005b, CQGra, 22, 3443 

\bibitem[Aurich, Lustig \& Steiner(2006)]{aurich:2006}
Aurich R., Lustig S., Steiner F., 2006, MNRAS, 369, 240

\bibitem[Aurich et al.(2007)]{aurich:2007} 
Aurich R., Lustig S., Steiner F., Then H., 2007, CQGra, 24, 1879

\bibitem[\protect\citeauthoryear{Aurich et al.}{2008}]{aurich:2008} 
Aurich R., Janzer H.~S., Lustig S., Steiner F., 2008, CQGra, 25, 125006 

\bibitem[\protect\citeauthoryear{Bennett et al.}{2003}]{bennett:2003}
Bennett C.~L., et al., 2003, ApJS, 148, 1

\bibitem[\protect\citeauthoryear{Bond et al.}{2000a}]{bond:2000a} 
Bond J.~R., Pogosyan D., Souradeep T., 2000a, Phys.~Rev.~D, 62, 043005 

\bibitem[\protect\citeauthoryear{Bond et al.}{2000b}]{bond:2000b} 
Bond J.~R., Pogosyan D., Souradeep T., 2000b, Phys.~Rev.~D, 62, 043006 

\bibitem[Caillerie et al.(2007)]{caillerie:2007}
Caillerie S., Lachieze-Rey M., Luminet J.~P., Lehoucq R., Riazuelo A., Weeks J.,
2007, A\&A, 476, 691

\bibitem[Copi et al.(2004)]{copi:2004} 
Copi C.~J., Huterer D., Starkman G.~D., 2004, Phys.~Rev.~D, 70, 043515

\bibitem[\protect\citeauthoryear{Copi et al.}{2006}]{copi:2006} 
Copi C.~J., Huterer D., Schwarz D.~J., Starkman G.~D., 2006, MNRAS, 367, 79

\bibitem[\protect\citeauthoryear{Cornish et al.}{1998}]{cornish:1998} 
Cornish N.~J., Spergel D.~N., Starkman G.~D., 1998, CQGra, 15, 2657 

\bibitem[\protect\citeauthoryear{Cornish et al.}{2004}]{cornish:2004} 
Cornish N.~J., Spergel D.~N., Starkman G.~D., Komatsu E., 2004, Phys.~Rev.~Lett., 92, 201302 

\bibitem[Dennis \& Land(2008)]{dennis:2008}
Dennis M.~R., Land K., 2008, MNRAS, 383, 424

\bibitem[de Oliveira-Costa et al.(2004)]{de Oliveira-Costa:2004}
de Oliveira-Costa A., Tegmark M., Zaldarriaga M., Hamilton A., 2004, Phys.~Rev.~D, 69, 063516

\bibitem[Eriksen et al.(2004)]{eriksen:2004} 
Eriksen H.~K., Hansen F.~K., Banday A.~J., G{\'o}rski K.~M., Lilje P. B., 2004, ApJ, 605, 14 

\bibitem[G{\'o}rski et al.(2005)]{healpix} 
G{\'o}rski K.~M., Hivon E., Banday A.~J., Wandelt B.~D., Hansen F.~K., Reinecke M., \& Bartelmann M., 2005, ApJ, 622, 759

\bibitem[\protect\citeauthoryear{Hajian \& Souradeep}{2003}]{hajian:2003} 
Hajian A., Souradeep T., 2003, preprint (astro-ph/0301590)

\bibitem[Hansen et al.(2004)]{hansen:2004} 
Hansen F.~K., Banday A.~J., G{\' o}rski K.~M., 2004, MNRAS, 354, 641

\bibitem[Hinshaw et al.(2007)]{hinshaw:2007}
Hinshaw G. et al., 2007, ApJS, 170, 288

\bibitem[Hinshaw et al.(2009)]{hinshaw:2009}
Hinshaw G. et.al., 2009, ApJS, 180, 225

\bibitem[\protect\citeauthoryear{Inoue}{2001}]{inoue:2001} 
Inoue K.~T., 2001, preprint (astro-ph/0103158)

\bibitem[Katz \& Weeks(2004)]{katz:2004}
Katz G., Weeks J., 2004, Phys.~Rev.~D, 70, 063527

\bibitem[\protect\citeauthoryear{Key et al.}{2007}]{key:2007} 
Key S.~J., Cornish N.~J., Spergel D.~N., Starkman G.~D., 2007, Phys.~Rev.~D, 75, 084034

\bibitem[\protect\citeauthoryear{Kunz et al.}{2006}]{kunz:2006} 
Kunz M., Aghanim N., Cayon L., Forni O., Riazuelo A., Uzan J.~P., 2006, Phys.~Rev.~D, 73, 023511 

\bibitem[\protect\citeauthoryear{Kunz et al.}{2008}]{kunz:2008} 
Kunz M., Aghanim N., Riazuelo A., Forni O., 2008, Phys.~Rev.~D, 77, 023525

\bibitem[Lachi\`eze-Rey(2004)]{mlr:2004}
Lachi\`eze-Rey M., preprint (astro-ph/0409081)

\bibitem[\protect\citeauthoryear{Land \& Magueijo}{2005}]{land:2005} 
Land K., Magueijo J., 2005, MNRAS, 362, L16

\bibitem[\protect\citeauthoryear{Levin et al.}{1998}]{levin:1998} 
Levin J., Scannapieco E., de Gasperis G., Silk J., Barrow J.~D., 1998, Phys.~Rev.~D, 58, 123006 

\bibitem[\protect\citeauthoryear{Levin}{2002}]{levin:2002} 
Levin J., 2002, Phys.~Rep., 365, 251 

\bibitem[\protect\citeauthoryear{Lew \& Roukema}{2008}]{lew:2008} 
Lew B., Roukema B., 2008, A\&A, 482, 747 

\bibitem[Luminet et al.(2003)]{luminet:2003}
Luminet J.-P., Weeks J., Riazuelo A., Lehoucq R., Uzan J.-P., 2003, Nature, 425, 593

\bibitem[Maxwell(1891)]{maxwell:1891}
Maxwell J.~C., 1891, A Treatise on Electricity and Magnetism, Clarendon Press

\bibitem[\protect\citeauthoryear{Niarchou \& Jaffe}{2006}]{niarchou:2006} 
Niarchou A., Jaffe A.~H., 2006, in Solomos N., Hellenic Naval Academy, eds, AIP Conf.~Proc. 848. Am.~Inst.~Phys., New York, p.~774 

\bibitem[\protect\citeauthoryear{Niarchou \& Jaffe}{2007}]{niarchou:2007} 
Niarchou A., Jaffe A.~H., 2007, Phys.~Rev.~Lett., 99, 081302

\bibitem[\protect\citeauthoryear{Riazuelo et al.}{2004a}]{riazuelo:2004a} 
Riazuelo A., Uzan J.-P., Lehoucq R., Weeks J., 2004a, Phys.~Rev.~D, 69, 103514

\bibitem[\protect\citeauthoryear{Riazuelo et al.}{2004b}]{riazuelo:2004b} 
Riazuelo A., Weeks J., Uzan J.-P., Lehoucq R., Luminet J.-P., 2004b, Phys.~Rev.~D, 69, 103518 

\bibitem[\protect\citeauthoryear{Roukema et al.}{2004}]{roukema:2004} 
Roukema B.~F., Lew B., Cechowska M., Marecki A., Bajtlik S., 2004, A\&A, 423, 821

\bibitem[\protect\citeauthoryear{Roukema et al.}{2007}]{roukema:2007} 
Roukema B.~F., Bajtlik S., Biesiada M., Szaniewska A., Jurkiewicz H., 2007, A\&A, 463, 861

\bibitem[\protect\citeauthoryear{Roukema et al.}{2008}]{roukema:2008} 
Roukema B.~F., Buli{\'n}ski Z., Szaniewska A., Gaudin N.~E., 2008, A\&A, 486, 55 

\bibitem[Schwarz et al.(2004)]{schwarz:2004} 
Schwarz D.~J., Starkman G.~D., Huterer D., Copi, C.~J., 2004, Phys.~Rev.~Lett., 93, 221301

\bibitem[\protect\citeauthoryear{Seljak \& Zaldarriaga}{1996}]{cmbfast}
Seljak U., Zaldarriaga M., 1996, ApJ, 469, 437

\bibitem[\protect\citeauthoryear{Tegmark, de Oliveira-Costa, \& Hamilton}{2003}]{tegmark:2003}
Tegmark M., de Oliveira-Costa A., Hamilton A.~J., 2003, Phys.~Rev.~D, 68, 123523

\bibitem[Weeks(2004)]{weeks:2004}
Weeks J.~R., 2004, preprint (astro-ph/0412231)

\bibitem[Wolf(1967)]{wolf:1967}
Wolf J., 1967, Space of constant curvature, McGraw-Hill, New York


\end{thebibliography}
\end{document}